\documentclass[aps, pra, 10pt, superscriptaddress, twocolumn, longbibliography, floatfix]{revtex4-1}

\usepackage[nointlimits]{amsmath}
\usepackage{amsthm}
\usepackage{graphicx,nicefrac}
\usepackage{subfigure}
\usepackage{color}
\usepackage{txfonts}
\usepackage{mathtools}
\usepackage{hyperref}
\usepackage{tabularx}
\hypersetup{colorlinks=true, citecolor=blue, linkcolor=blue}

\usepackage{amsfonts,amsmath,amssymb}
\usepackage{array,bm,color}
\usepackage{epsfig,graphicx,nomencl,revsymb4-1,upgreek,url}
\usepackage{hyperref}
\usepackage{algpseudocode}
\usepackage{xspace}
\usepackage{tabularx}
\usepackage[normalem]{ulem}
\usepackage{changes}
\usepackage{enumitem}
\usepackage{booktabs}
\usepackage{mathtools}

\def\bra#1{{\left\langle #1 \right|}}
\def\ket#1{{\left| #1 \right\rangle}}

\newcommand{\pauli}{\mathbb{P}}
 
\newcommand{\Id}{I}

\DeclareMathOperator*{\E}{\mathbb{E}}
\newcommand{\Tr}{\text{Tr}}

\newcommand{\Q}[1]{\texttt{Q#1}}

\newcommand{\cnot}{\textsc{cnot}}

\begin{document}
\title{Fully scalable randomized benchmarking without motion reversal}
\author{Jordan Hines}
\thanks{jordanh@berkeley.edu}
\affiliation{Department of 
Physics, University of California, Berkeley, CA 94720}
\affiliation{Quantum Performance Laboratory, Sandia National Laboratories, Albuquerque, NM 87185 and Livermore, CA 94550}
\author{Daniel Hothem}
\author{Robin Blume-Kohout}
\affiliation{Quantum Performance Laboratory, Sandia National Laboratories, Albuquerque, NM 87185 and Livermore, CA 94550}
\author{Birgitta Whaley}
\affiliation{Department of Chemistry, University of California, Berkeley, CA 94720}
\author{Timothy Proctor}
\thanks{tjproct@sandia.gov}
\affiliation{Quantum Performance Laboratory, Sandia National Laboratories, Albuquerque, NM 87185 and Livermore, CA 94550}
\begin{abstract} 
We introduce \emph{binary randomized benchmarking (BiRB)}, a protocol that streamlines traditional RB by using circuits consisting almost entirely of i.i.d.~layers of gates. BiRB reliably and efficiently extracts the average error rate of a Clifford gate set by sending tensor product eigenstates of random Pauli operators through random circuits with i.i.d.~layers. Unlike existing RB methods, BiRB does not use motion reversal circuits---i.e., circuits that implement the identity (or a Pauli) operator---which simplifies both the method and the theory proving its reliability. Furthermore, this  simplicity enables scaling BiRB to many more qubits than the most widely-used RB methods. 

\end{abstract}
\maketitle
\section{Introduction}

\begin{figure}[ht]
    \centering
    \includegraphics{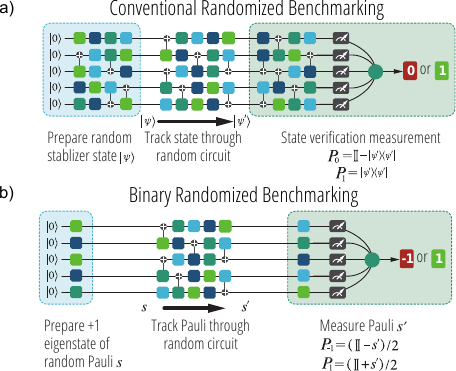}
    \caption{\textbf{RB without motion reversal.} (a) Standard RB methods use motion reversal circuits, which make errors easily visible but add complexity and limit scalability. 
    (b) BiRB reliably estimates average gate error rates without motion reversal by tracking a single stabilizer of a random product state through a random circuit. 
    (c) Results from CRB, DRB, and BiRB on \texttt{ibm\_hanoi} show that BiRB is more scalable than both CRB and DRB.
    DRB estimates the same error rate as BiRB ($r_{\Omega}$), and we find that their error rates are consistent, providing evidence for the reliability of BiRB.}
    \label{fig:fig1}
\end{figure}

Randomized benchmarking (RB) \cite{proctor2021scalable, proctor2020measuring,  McKay2020-no, emerson2005scalable, emerson2007symmetrized, magesan2011scalable, magesan2012characterizing, knill2008randomized, carignan2015characterizing, cross2016scalable, brown2018randomized, hashagen2018real, magesan2011scalable, magesan2012characterizing,  carignan2015characterizing, cross2016scalable, brown2018randomized, hashagen2018real, helsen2018new, Helsen2020-it, Claes2020-cy,  Helsen2020-mb, Morvan2020-ck, proctor2018direct, cross2018validating, Mayer-df, hines2022demonstrating} is a family of protocols that assess the average performance of a quantum processor's gates by running random circuits. RB experiments are ubiquitous, yet the most widely-used RB protocols have important limitations that are caused by the kind of random circuits they use. Most RB protocols use \emph{motion reversal} circuits that, if run without errors, implement the identity (or a Pauli) operator \cite{emerson2005scalable, magesan2011scalable, emerson2007symmetrized, proctor2018direct,proctor2021scalable, carignan2015characterizing, cross2016scalable, helsen2022framework} [Fig.~\ref{fig:fig1}(a)]. This makes errors easily visible: each RB circuit, when run perfectly, always outputs a particular bit string, so the observation of any other bit string implies that an error occurred. However, random motion reversal circuits must end with an \emph{inversion subroutine} that undoes the preceding layers. The inversion subroutine causes  challenges for RB theory \cite{wallman2017randomized, polloreno2023direct, proctor2017randomized, magesan2011scalable, magesan2012characterizing, merkel2018randomized, proctor2021scalable, hines2022demonstrating} as well as practical problems. In most existing RB techniques---including standard Clifford group RB (CRB) \cite{magesan2011scalable} and its streamlined variant direct RB (DRB) \cite{proctor2018direct}---the size of the inversion subroutine grows quickly with the number of qubits \cite{aaronson2004improved, patel2003efficient, bravyi2020hadamard} [see Fig.~\ref{fig:fig1}(a)], severely limiting their applicability outside of the few-qubit setting \cite{proctor2018direct, McKay2020-no, proctor2021scalable}.

\begin{figure}[h]
    \centering
    \includegraphics{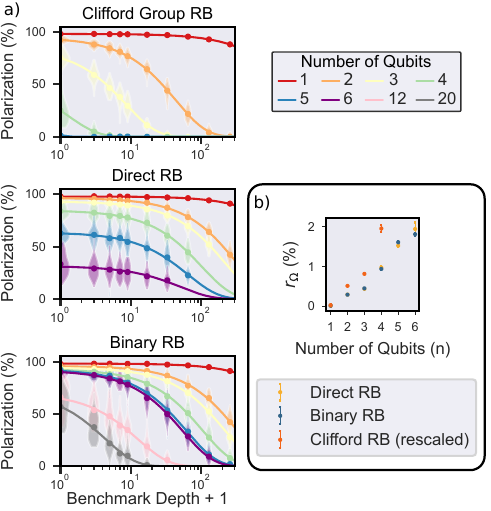}
    \caption{\textbf{BiRB and conventional RB on IBM Q.} Results from CRB, DRB, and BiRB on \texttt{ibm\_hanoi} show that BiRB is more scalable than both CRB and DRB.
    DRB estimates the same error rate as BiRB ($r_{\Omega}$), and we find that their error rates are consistent, providing evidence for the reliability of BiRB.}
    \label{fig:fig2}
\end{figure}

In this work, we demonstrate that motion reversal circuits are not required for reliable RB by introducing \emph{binary randomized benchmarking (BiRB)}. BiRB is an efficient and scalable protocol for estimating the average error rate of a Clifford gate set. BiRB's circuits [Fig.~\ref{fig:fig1}(b)] consist of $d$ i.i.d.~layers of gates and two layers of single-qubit gates, for state and measurement preparation, and the measurement results are processed to obtain a binary-outcome Pauli measurement result.
BiRB works because the average fidelity of highly scrambling random circuits decays exponentially in depth \cite{proctor2018direct, polloreno2023direct, hines2022demonstrating} and, for Clifford circuits, this fidelity can be efficiently
estimated using random local state preparations and measurements \cite{Flammia_2011, da_Silva_2011}.
Our method's local state preparation and measurement enables benchmarking of many more qubits than most existing RB techniques---including CRB and DRB---as shown in Fig.~\ref{fig:fig2}.
Furthermore, we show that BiRB is more accurate than mirror RB (MRB) \cite{proctor2021scalable, hines2022demonstrating}, which is the only other scalable RB protocol for Clifford gate sets.

BiRB connects RB and cross entropy benchmarking (XEB) \cite{boixo2018characterizing, arute2019quantum, liu2021benchmarking, chen2022linear}, another form of randomized benchmark. In contrast to RB, XEB uses random circuits consisting solely of i.i.d.~(composite) layers of gates---these layers are typically sampled from a universal gate set, but a scalable form of XEB using Clifford gate sets has also been introduced \cite{chen2022linear}. While XEB circuits have no overhead from subroutines, in practice XEB decay curves exhibit non-exponential behavior at low depths for some Markovian error models, and therefore measuring a reliable error rate requires circuits with at least $O(n)$ depth \cite{heinrich2023randomized}.  The exact circuit depths required for exponential decay depend on the connectivity and gate set and must be estimated numerically for each distribution of layers benchmarked, adding additional complication to performing XEB. This issue arises in part because XEB estimates the fidelity of random circuits using the (linear) cross entropy, which is not an accurate fidelity estimator for general Markovian noise models \cite{gao2021limitations, liu2021benchmarking}. BiRB shows how to add minimal overhead to circuits of i.i.d.~layers to obtain a provably reliable RB protocol. 

The remainder of this paper is structured as follows. In Section~\ref{sec:prelim} we introduce our notation and review the existing results on which our method relies. In Section~\ref{sec:method} we introduce the BiRB protocol. In Section~\ref{sec:theory} we present a theory of BiRB that shows that our method is reliable: it accurately estimates the average error rate of an $n$-qubit circuit layer under assumptions commonly used in RB theory (e.g., Markovian errors). In Section~\ref{sec:simulations} we demonstrate the reliability of our method with numerical simulations of BiRB on gate sets that experience both stochastic Pauli errors and (coherent) Hamiltonian errors. In Section~\ref{sec:demos}, we demonstrate BiRB on IBM Q processors and validate it against the results of DRB and MRB. We then conclude in Section~\ref{sec:discussion}.

\section{Preliminaries}
\label{sec:prelim}
\subsection{Definitions}
In this section, we introduce our notation. An $n$-qubit \emph{layer} $L$ is an instruction to perform a particular unitary operation on those $n$ qubits, typically specified in terms of 1- and 2-qubit gates. We use $U(L) \in \textrm{SU}(2^n)$ to denote the unitary corresponding to $L$. The layers we use are randomly sampled, and we often treat a layer $L$ as a layer-valued random variable. We use $\Omega: \mathbb{L} \rightarrow [0,1]$ to denote a probability distribution over the set of layers $\mathbb{L}$. We use $L^{-1}$ to denote an instruction to perform the unitary $U(L)^{-1}$. An $n$-qubit, depth-$d$ \textit{circuit} is a sequence of $n$-qubit layers $C = L_dL_{d-1}\cdots L_{2} L_1$, where we use the convention that the circuit is read right to left. 

For a layer (or circuit) $L$, we use $\mathcal{U}(L)$ to denote the superoperator representation of its perfect implementation, i.e., $\mathcal{U}(L)[\rho] = U(L) \rho U^{\dagger}(L)$. We use $\phi(L)$ to denote the superoperator for an imperfect implementation of $L$, and we assume $\phi(L)$ is a completely positive trace preserving (CPTP) map. A layer $L$'s \emph{error map} is defined by $\mathcal{E}_L = \phi(L)\mathcal{U}^{\dagger}(L)$.  The \emph{entanglement fidelity} (also called the \emph{process fidelity}) of $\phi(L)$ to $\mathcal{U}(L)$ is defined by
\begin{align}
    F\bigl(\phi(L), \mathcal{U}(L)\bigr) = F(\mathcal{E}_L) & = \langle \varphi | \bigl(\mathbb{I} \otimes \mathcal{E}_L\bigr)[|\varphi \rangle \langle \varphi |]|\varphi \rangle \\
    & = \frac{1}{4^n}\Tr(\mathcal{U}(L)^{\dag}\phi(L)) \\
    & = \E\limits_{s \in \pauli_n} \Tr(s\mathcal{E}_L[s]), \label{eq:paulis_tr}
\end{align}
where $\varphi$ is any maximally entangled state of $2n$ qubits \cite{nielsen2002simple}, and $\pauli_n$ is the set of all $n$-qubit Pauli operations with $\pm 1$ global sign. Throughout, we use the term ``(in)fidelity'' to refer to the entanglement (in)fidelity. The \emph{polarization} is a rescaling of fidelity given by 
\begin{align}
    \gamma(\phi(L), \mathcal{U}(L)) = \gamma(\mathcal{E}_L) & = \frac{4^n}{4^n-1} F(\mathcal{E}_L) -  \frac{1}{4^n-1} \label{eq:pol_def} \\
      &  = \E\limits_{s \in \pauli^{\ast}_n} \Tr(s\mathcal{E}_L[s]), \label{eqn:est_pol}
\end{align}
where $\pauli^{\ast}_n = \pauli_n \setminus \{\pm \Id_n\}$, and $\Id_n$ denotes the $n$-qubit identity operator.  We say that a state $\ket{\psi}$ is \emph{stabilized} by a Pauli operator $P$ if $P\ket{\psi}=\ket{\psi}$. An $n$-qubit \emph{stabilizer state} $\ket{\psi}$ is a state that is stabilized by exactly $2^n$ Pauli operators. Equivalently, a stabilizer state is a state that can be prepared from $\ket{0}^{\otimes n}$ using only Clifford gates \cite{aaronson2004improved}. The \emph{stabilizer group} of a stabilizer state $\ket{\psi}$ is $S_{\psi} = \{P \in \mathbb{P}_n \mid P\ket{\psi} = \ket{\psi}\}$. We use $S^{\ast}_{\psi}$ to denote all non-identity elements of the stabilizer group, i.e., $S^{\ast}_{\psi} = S_{\psi} \setminus \{\Id_n\}$.

\subsection{$\Omega$-distributed random circuits}

BiRB uses \emph{$\Omega$-distributed random circuits} \cite{proctor2021scalable, hines2022demonstrating, proctor2018direct, polloreno2023direct}, which we now review. $\Omega$-distributed random circuits consist of $n$-qubit layers of gates sampled from a distribution $\Omega(\mathbb{L})$ over a layer set $\mathbb{L}$. In this work, we restrict $\mathbb{L}$ to contain only Clifford gates. These circuit layers can be chosen to consist of a processor’s native gates, or simple combinations thereof, thus eliminating the need for complicated compilation. 

$\Omega$-distributed random circuits are also used in DRB and MRB \cite{proctor2021scalable, hines2022demonstrating, proctor2018direct, polloreno2023direct}. DRB and MRB are reliable if $\Omega$ satisfies certain conditions, and these same conditions are required for BiRB to be reliable. We require that the circuits generated by layers sampled from $\Omega$ are highly scrambling, meaning that for all Pauli operators $P, P' \neq \Id_n$, there exists constants $k \ll \nicefrac{1}{\varepsilon}$ and $\delta \ll 1$ such that
    \begin{equation}
        \frac{1}{4^n}\E\limits_{L_1, \cdots, L_k} Tr(\mathcal{P}' \mathcal{U}(L_k\cdots L_1)\mathcal{P} \mathcal{U}(L_k \cdots L_1)^{-1}) \leq \delta + \frac{1}{4^n}.\label{eqn:scrambling}
    \end{equation}
Here, $\mathcal{P}[\rho] = P\rho P$ and $\mathcal{P}'[\rho] = P'\rho P'$ are Pauli superoperators, $L_1, \dots, L_k$ are $\Omega$-distributed random layers, and $\varepsilon$ is the expected infidelity of an $\Omega$-distributed random layer \cite{hines2022demonstrating, polloreno2023direct}. Informally, this condition means that an error is locally randomized (i.e., its basis is randomized over the $X$, $Y$, and $Z$ bases) and delocalized across multiple qubits before a second error is likely to have occurred. While we require that $k \ll \nicefrac{1}{\varepsilon}$ for our theory, this condition on $k$ can be relaxed when $n \gg 1$, because errors that occur on spatially separated qubits in close succession cannot cancel at all (see Refs.~\cite{hines2022demonstrating, polloreno2023direct} for details).

\subsection{The RB error rate}
\label{sec:epsilon_omega}
BiRB's output is an error rate $r_{\Omega}$ that quantifies the error in random $n$-qubit layers sampled from $\Omega$. BiRB's $r_{\Omega}$ closely approximates an independent, physically motivated error rate $\epsilon_{\Omega}$---which is closely related to the average layer infidelity---introduced in Refs.~\cite{carignan2018randomized, hines2022demonstrating} and reviewed here. $\epsilon_{\Omega}$ is defined by the \emph{rate of decay} of the fidelity of $\Omega$-distributed random circuits. The expected fidelity of depth-$d$ $\Omega$-distributed random circuits $C_d$ is given by 
\begin{equation}
    \bar{F}_d = \E\limits_{C_d}
    F\bigl(\phi(C_d),\mathcal{U}(C_d)\bigr). \label{eq:f_d_def}
\end{equation}
The scrambling requirements on $\Omega$ (see Section~\ref{sec:circuits}) ensure that $\bar{F}_d$ [\ref{eq:f_d_def}] decays exponentially, i.e., $\bar{F}_d \approx Ap_{\textrm{rc}}^d + B$ for constants $A, B,$ and $p_{\textrm{rc}}$. The average error rate of layers sampled from $\Omega$ is then defined as \cite{carignan2018randomized, hines2022demonstrating}
\begin{equation}
    \epsilon_{\Omega} = \frac{4^n-1}{4^n}(1-p_{\textrm{rc}}). \label{eqn:epsilon_Omega}
\end{equation}
This rescaling of $p_{\textrm{rc}}$ is used because $p_{\textrm{rc}}$ corresponds to the effective polarization of a random layer in an $\Omega$-distributed random circuit---i.e., the polarization in a depolarizing channel that would give the same fidelity decay---so $\epsilon_{\Omega}$ is the effective average infidelity of a layer sampled from $\Omega$. When stochastic Pauli errors are the dominant source of error, $\epsilon_{\Omega}$ is approximately equal to the average layer infidelity, 
\begin{equation}
\varepsilon_{\Omega} = 1 - \E_{L \in \mathbb{L}} F(\phi(L), \mathcal{U}(L)),
\end{equation}
but this is not true more generally because gate infidelity is not ``gauge-invariant''---see Refs.~\cite{carignan2018randomized, hines2022demonstrating, proctor2017randomized, wallman2017randomized} for details.

\subsection{Direct fidelity estimation}
\label{sec:method_intro}
Our protocol can be interpreted as an application of \emph{direct fidelity estimation} (DFE) \cite{Flammia_2011, da_Silva_2011} to varied-depth random Clifford circuits, so we now review DFE for the special case of Clifford circuits.  
Consider a Clifford circuit $C$ and an imperfect implementation of that circuit $\phi(C) =\mathcal{U}(C)\mathcal{E}_C$, where $\mathcal{E}_C$ denotes the overall error map of the circuit. 
Using Eq.~\eqref{eqn:est_pol}, the polarization of $\mathcal{E}_C$ can be written as
\begin{align}
    \gamma(\mathcal{E}_C) 
    & = \E\limits_{s \in \pauli^{\ast}_n} \Tr(s\mathcal{E}_C[s]) \\
    & = \E\limits_{s \in \pauli^{\ast}_n} \Tr(s\mathcal{U}(C)^{\dag}\phi(C)[s]) \\
    &= \E\limits_{s \in \pauli^{\ast}_n} \Tr(s' \mathcal{\phi}(C)[s]), 
    \label{eqn:est_pol_non_id}
\end{align}
where $s' = U(C)sU(C)^{\dag}$ is a Pauli operator that can be efficiently computed classically \cite{aaronson2004improved}, because $C$ is a Clifford circuit. Eq.~\eqref{eqn:est_pol_non_id} implies that polarization of $\mathcal{E}_C$ can be efficiently estimated as follows: (1) sample Pauli operators uniformly from $\mathbb{P}_n^{\ast}$, (2) for each sampled Pauli operator $s$, apply $\phi(C)$ to $s$ and measure the evolved Pauli operator $s'$, and (3) average the measurement results. It is not physically possible to directly apply $\phi(C)$ to a Pauli operator $s$ (Pauli operators are not valid quantum states), but DFE simulates doing so by applying $\phi(C)$ to randomly sampled eigenstates of $s$. 
BiRB also uses this approach, but, unlike DFE, BiRB is robust to state preparation and measurement (SPAM) error. BiRB separates SPAM error from gate error by applying DFE to variable-depth circuits and extracting gate error from the rate of decay of the polarization---as in cycle benchmarking \cite{erhard2019characterizing} and Pauli noise learning techniques \cite{flammia2021averaged, harper2019efficient,  flammia2019efficient}.

\section{The binary RB protocol}
\label{sec:method}
We now introduce BiRB circuits (Section~\ref{sec:circuits}) and the BiRB protocol (Section~\ref{sec:protocol}). 

\begin{figure}[t!]
    \centering    
    \includegraphics{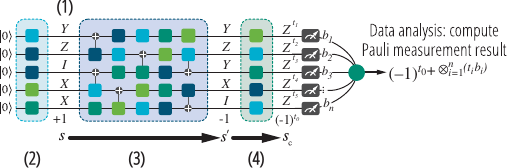}
    \caption{\textbf{BiRB circuits.} Each BiRB circuit is constructed by (1) generating a random Pauli $s$, then constructing a circuit consisting of (2) a layer of single-qubit gates generating a +1 eigenstate of a random Pauli operator $s$, (3) $d$ layers of Clifford gates randomly sampled from some distribution $\Omega$, and (4) a final layer of gates that transforms the evolved Pauli operator ($s'$) into a tensor product of $Z$ and $I$ Pauli operators ($s_C$, represented by bit string $t$). The result of a computational basis measurement (bit string $b$) is used to compute a 1 (``success'') or $-1$ (``fail'') result by comparing it to the bit string $t$.}
    \label{fig:circuit}
\end{figure}

\subsection{Binary RB circuits}
\label{sec:circuits}
We now state the procedure for constructing BiRB circuits [Fig.~\ref{fig:circuit}(b)]. Each BiRB circuit first generates an eigenstate of a random Pauli operator $s$, then applies a depth $d$ random circuit, and then ends with a measurement of the evolved Pauli operator $s'$. A width $n$, benchmark depth $d$, $\Omega$-distributed BiRB circuit is a circuit $C = L_{d+1}L_{d}\cdots L_{1}L_{0}$ that begins with preparing $\ket{0}^{\otimes n}$ and ends with a computational basis measurement, and has layers sampled as follows:

\begin{enumerate}
    \item Sample a uniformly random $n$-qubit Pauli operator $s \in \pauli_n^{\ast}$ and a uniformly random state $\ket{\psi(s)}$ from the set of tensor product stabilizer states stabilized by $s$. $L_0$ is a layer of single-qubit gates that prepares $\ket{\psi(s)}$. 
    \item $L_1, L_2, \dots, L_d$ are layers sampled from $\Omega$. These layers form the \emph{core circuit}, which is a depth-$d$ $\Omega$-distributed random circuit.
    \item\label{last_layer} $L_{d+1}$ is a layer of single-qubit gates that transforms
    \begin{equation}
        s'=U(L_d)\cdots U(L_1)sU(L_1)^{-1}\cdots U(L_d)^{-1}
    \end{equation} 
     into a tensor product of $Z$ and $I$ operators.  
\end{enumerate}
The circuit has an associated ``target" Pauli operator
        \begin{equation}
        s_C =U(L_{d+1})s'U(L_{d+1})^{-1}. \label{eqn:s_c}
    \end{equation} 
If implemented without errors, the bit string $b$ output by $C$ will correspond to a $+1$ eigenstate of $s_C$, i.e., $s_C\ket{b}=\ket{b}$. 

Step (1) can equivalently be formulated as (i) sampling a random unsigned Pauli $P$, (ii) picking a random tensor product stabilizer state that is an eigenstate of $P$. Sampling from both $+1$ and $-1$ eigenstates ensures accurate fidelity estimation when there are non-unital errors in the circuits (see Section~\ref{sec:theory}).

There is not a unique choice for either the initial layer ($L_0$) or the final layer ($L_{d+1}$) of gates in BiRB circuits. These layers may be chosen deterministically or at random from the set of all possible layers of single-qubit Clifford gates satisfying the criteria above. In our simulations and experiments, we choose to randomize $L_{0}$, but this is not required for our theory. There will always be a possible final layer satisfying the requirements in step~\ref{last_layer}. We can construct such a layer as follows: Let $s' = \bigotimes_{i=1}^n s'_i$, where $s'_i$ denotes the single-qubit Pauli operator acting on qubit $i$. On qubit $i$, apply $H$ if $s'_i=X$, apply $HS^{\dag}$ if $s'_i=Y$, and apply $I$ if $s'_i=I$ or $Z$.

\subsection{Binary RB protocol}
\label{sec:protocol}
The BiRB protocol is similar to other RB protocols: run BiRB circuits, compute a figure of merit for the circuits of each benchmark depth, then fit an exponential decay. 
A BiRB experiment is defined by a layer set $\mathbb{L}$, a sampling distribution $\Omega$, and the usual RB sampling parameters (a set of benchmark depths $d$, the number of circuits $K$ sampled per depth, and the number of times $N$ each circuit is run). Our protocol is the following:
\begin{enumerate}
    \item For a range of integers $d \geq 0$, sample $K$ $\Omega$-distributed BiRB circuits with benchmark depth $d$, and run each circuit $N\geq 1$ times. 
    \item For each circuit $C$, estimate the expected value $\langle s_C \rangle$ of the target Pauli observable $s_C$ from the computational basis measurement results. Then, compute the average over all circuits of benchmark depth $d$, 
    \begin{equation}
        \bar{f}_d = \frac{1}{K}\sum\limits_{C_d} \langle s_C \rangle. \label{eqn:fbar_d}
    \end{equation}
    \item Fit $\bar{f}_d$ to an exponential, 
    \begin{equation}
        \bar{f}_d = Ap^d,
    \end{equation} where $A$ and $p$ are fit parameters. Then compute \footnote{We choose this definition of $r_{\Omega}$ so that it corresponds to the average layer entanglement infidelity. It is also acceptable to set $r_{\Omega} = (2^n - 1)(1 - p)/2^n$, so that it corresponds to average gate infidelity.}
   \begin{equation}
   r_{\Omega} = (4^n - 1)(1 - p)/4^n . \label{eqn:r_Omega}
   \end{equation}
\end{enumerate}

In Appendix~\ref{app:statistics}, we show that the number of circuits per depth (and therefore the total amount of data) required by our protocol to estimate $r_{\Omega}$ to within a fixed relative uncertainty  is independent of $n$. For a fixed total number of shots per depth $KN$, it is statistically optimal to maximize the number of random circuits $K$ and set $N=1$. However, typically $N>1$ for practical reasons---e.g., because of the time cost of generating and loading many distinct circuits onto the processor. See Refs.~\cite{harper2019statistical, wallman2014randomized, kwiatkowski2023optimized} for more detailed statistical analyses of RB protocols. 

Note that if $\mathbb{L}$ is chosen to be the set of all $n$-qubit Clifford gates and $\Omega$ is the uniform distribution, then we obtain a version of standard RB (i.e., RB of the Clifford group) without an inversion gate. See Appendix~\ref{app:standard_rb} for further discussion of this variant of BiRB, whose reliability can be proven using the unitary 2-design twirling theory that underpins the theory of standard RB \cite{magesan2011scalable, magesan2012characterizing}. 

\section{Theory of Binary RB}
\label{sec:theory}
We now show that the error rate measured by BiRB [$r_{\Omega}$, Eq.~\eqref{eqn:r_Omega}] is a close approximation to the average layer error rate $\epsilon_{\Omega}$ [Eq.~\eqref{eqn:epsilon_Omega}]. In Section~\ref{sec:l_op} we show that BiRB estimates the expected fidelity of depth-$d$ $\Omega$-distributed circuits. In Section~\ref{sec:fidelity_decay}, we show that this quantity decays exponentially in $d$, which allows us to conclude that the BiRB error rate is approximately the average layer error rate, i.e., $r_{\Omega} \approx \epsilon_{\Omega}$.

\subsection{Relating measurement results to circuit polarizations}
\label{sec:l_op}
We start by showing that $\bar{f}_d$ [Eq.~\eqref{eqn:fbar_d}] is approximately equal to the expected polarization [Eq.~\eqref{eq:pol_def}] of an error map consisting of the composition of (1) the error map of a depth-$d$ $\Omega$-distributed random circuit, and (2) an error map absorbing all state preparation and measurement (SPAM) error. We then argue that the contribution of SPAM errors is approximately depth-independent and can be factored out, so that $\bar{f}_d$ equals the polarization [Eq.~\eqref{eq:pol_def}] of the error map of a random, depth-$d$ $\Omega$-distributed random circuit, multiplied by a $d$-independent prefactor.

We consider a BiRB circuit $C$ with benchmark depth $d$ and gate-dependent error channels on $L_1, L_2, \dots, L_d$, i.e., $\phi(L)=\mathcal{E}_L\mathcal{U}(L)$. We model the error on $L_0$ and state preparation as a gate-independent global depolarizing channel $\mathcal{E}_0$ directly after $L_0$. We model the error on $L_{d+1}$ and readout as a single, gate- and measurement-independent global depolarizing channel $\mathcal{E}_{d+1}$ occurring directly before $L_{d+1}$. Therefore, the superoperator representing the imperfect implementation of the circuit $C$ is given by 
\begin{equation}
    	\phi(C) = \mathcal{U}(L_{d+1}) \mathcal{E}_{{d+1}} \mathcal{E}_{L_d}\mathcal{U}(L_d) \cdots \mathcal{E}_{L_1}\mathcal{U}(L_1) \mathcal{E}_{0}\mathcal{U}(L_0). \label{eqn:phi_c}
\end{equation}

We first rewrite the error in the circuit in terms of the core circuit's error map.  We have
\begin{equation}
	\phi(C)   =  \mathcal{U}(L_{d+1} L_d \cdots L_1) \mathcal{E}_{\textrm{tot}} \mathcal{U}(L_0) ,\label{eqn:l_op_phi_expansion}
\end{equation}
where $\mathcal{U}(L_{d+1} L_d \cdots L_1) =  \mathcal{U}(L_{d+1})\mathcal{U}(L_d) \cdots \mathcal{U}(L_1)$, and $\mathcal{E}_{\textrm{tot}} = \mathcal{E}_{d+1}\mathcal{E}_{L_1, \cdots, L_d} \mathcal{E}_0$ for 
\begin{equation}
    \mathcal{E}_{L_1, \cdots L_d}  = \mathcal{U}(L_1)^{-1}\cdots \mathcal{U}(L_d)^{-1}\mathcal{E}_{L_d}\mathcal{U}(L_d)\cdots\mathcal{E}_{L_1}\mathcal{U}(L_1). \label{eqn:l_op_channel} 
\end{equation}
Now we show that $\bar{f}_d$ is the expected polarization of $\mathcal{E}_{\textrm{tot}}$. $\bar{f}_d$ is the expectation value of circuit $C$'s target Pauli operator $s_C$ [Eq.~\eqref{eqn:s_c}] averaged over all benchmark depth-$d$ BiRB circuits, i.e.,
\begin{equation}
    \bar{f}_d  = \E_{L_1,\dots,L_d}\E_{s \in \pauli_n^{\ast}}\E_{\ket{\psi(s)}}\Tr(s_C\mathcal{U}(L_{d+1} \cdots L_{1}) \mathcal{E}_{\textrm{tot}}[\ket{\psi(s)}\bra{\psi(s)}]),\label{eqn:l_op_f_unsimplified} 
\end{equation}
where $\ket{\psi(s)} = U(L_0)\ket{0}^{\otimes n}$ is a uniformly random state from the set of all tensor product states stabilized by $s$. Substituting in $s_c = \mathcal{U}(L_{d+1}\cdots L_{1})[s]$, Eq.~\eqref{eqn:l_op_f_unsimplified} becomes
\begin{equation}
\bar{f}_d  =\E_{L_1,\dots,L_d}\E_{s \in \pauli_n^{\ast}} \E_{\ket{\psi(s)}} \Tr(s \mathcal{E}_{\textrm{tot}}[\ket{\psi(s)}\bra{\psi(s)}]). \label{eqn:l_op_tr_error_channel}
\end{equation}

We now average over $\ket{\psi(s)}$. To do so, we first expand the initial state $\ket{\psi(s)}$ in terms of its stabilizer group:
\begin{align}
    \bar{f}_d & = \frac{1}{2^n} \E_{L_1,\dots,L_d} \E_{s \in \pauli_n^{\ast}} \E_{\ket{\psi(s)}} \Tr\left(s \mathcal{E}_{\textrm{tot}}\left[\sum_{s' \in S_{\ket{\psi(s)}}}s'\right]\right) \\
    & = \frac{1}{2^n} \E_{L_1,\dots,L_d} \E_{s \in \pauli_n^{\ast}} \Tr\left(s \mathcal{E}_{\textrm{tot}}\left[\Id_n + s +  \E_{\ket{\psi(s)}} \sum_{\substack{s' \in S_{\ket{\psi(s)}}\\ s' \neq \Id_n , s}}s'\right]\right) \label{eqn:f_stabilizer_expansion2} \\
    & = \frac{1}{2^n}\E_{L_1,\dots,L_d} \E_{s \in \pauli_n^{\ast}} \left(\Tr(s \mathcal{E}_{\textrm{tot}}[s]) +  \Tr\left(s \E_{\ket{\psi(s)}} \sum_{\substack{s' \in S_{\ket{\psi(s)}}\\ s' \neq \Id_n , s}} \mathcal{E}_{\textrm{tot}}[s']\right)\right). \label{eqn:f_stabilizer_expansion}
\end{align}
To get from Eq.~\eqref{eqn:f_stabilizer_expansion2} to Eq.~\eqref{eqn:f_stabilizer_expansion}, we use the fact that $\E_{s \in \pauli_n^{\ast}} \Tr(s\mathcal{E}_{\textrm{tot}}[I_n])=0$ because we are averaging over signed non-identity Pauli operators. The symmetry properties of the set of all local $+1$ eigenstates of $s$ guarantee that the last term of Eq.~\eqref{eqn:f_stabilizer_expansion} vanishes (see Appendix~\ref{app:theory}), so that Eq.~\eqref{eqn:f_stabilizer_expansion} becomes
\begin{align}
    \bar{f}_d & = \frac{1}{2^n} \E_{L_1,\dots,L_d}\E_{s \in \pauli_n^{\ast}} \Tr(s \mathcal{E}_{\textrm{tot}}[s]) \label{eqn:f_d_simple} \\
    & =\E_{L_1,\dots,L_d}  \gamma(\mathcal{E}_{\textrm{tot}})\label{eqn:f_equals_pol}.
\end{align}

Eq.~\eqref{eqn:f_equals_pol}  says that $\bar{f}_d$, which is measured in our protocol, is the expected polarization of $\mathcal{E}_{\textrm{tot}}$. This error map is the composition of (1) the error map of an $\Omega$-distributed random circuit and (2) the error maps of the state preparation and measurement layers. Because $\mathcal{E}_0$ and $\mathcal{E}_{d+1}$ are (by assumption) global depolarizing channels, we have
\begin{equation}
    \bar{f}_d =  \gamma(\mathcal{E}_0)\gamma(\mathcal{E}_{d+1}) \E_{L_1,\dots,L_d} \gamma(\mathcal{E}_{L_1, \dots, L_d}).  \label{eqn:f_d_result}
\end{equation}

If $\mathcal{E}_0$ and $\mathcal{E}_{d+1}$ are stochastic Pauli channels (but not necessarily global depolarizing channels), or if $\mathcal{E}_{L_1, \dots, L_d}$ is a stochastic Pauli channel, then Eq.~\eqref{eqn:f_d_result} holds approximately. Specifically,
\begin{equation}
    \bar{f}_d =  \gamma(\mathcal{E}_0\mathcal{E}_{d+1}) \E_{L_1,\dots,L_d} \gamma(\mathcal{E}_{L_1, \dots, L_d}) + O(\varepsilon_{\textrm{SPAM}}\varepsilon_{L_1, \dots, L_d}),  \label{eqn:f_d_result_2}
\end{equation}
where $\varepsilon_{\textrm{SPAM}}$ is the infidelity of $\mathcal{E}_0\mathcal{E}_{d+1}$ and $\varepsilon_{L_1, \dots, L_d}$ is the infidelity of $\mathcal{E}_{L_1, \dots, L_d}$ \footnote{If $\mathcal{E}_0\mathcal{E}_{d+1}$ is a stochastic Pauli channel, then $\gamma(\mathcal{E}_0\mathcal{E}_{d+1}\mathcal{E}_{L_1, \dots, L_d}) = \E_{P \in \pauli_n} \gamma(\mathcal{P}\mathcal{E}_0\mathcal{E}_{d+1}\mathcal{P}^{\dag}\mathcal{E}_{L_1, \dots, L_d}) = \gamma(\mathcal{E}_0\mathcal{E}_{d+1}(\E_{P \in \pauli_n}\mathcal{P}^{\dag}\mathcal{E}_{L_1, \dots, L_d}\mathcal{P}))$. Eq.~\eqref{eqn:f_d_result_2} then follows because $\mathcal{E}_0\mathcal{E}_{d+1}$ and $\E_{P \in \pauli_n}\mathcal{P}^{\dag}\mathcal{E}_{L_1, \dots, L_d}\mathcal{P}$ are stochastic Pauli channels \cite{proctor2021scalable}.}. The size of the $O(\varepsilon_{\textrm{SPAM}}\varepsilon_{L_1, \dots, L_d})$ term is determined by the amount of error cancellation between $\mathcal{E}_0\mathcal{E}_{d+1}$ and $\mathcal{E}_{L_1,\dots,L_d}$ \cite{proctor2021scalable}. At low depths $d$, this correction term is small  because $\varepsilon_{L_1, \dots, L_d}$ is small, and at depths $d \gtrsim k$ [where $k$ is the small constant in Eq.~\eqref{eqn:scrambling}], this term is small because the scrambling condition for $\Omega$-distributed random layers implies that errors in that circuit are randomized and spread over many qubits. Eq.~\eqref{eqn:f_d_result_2} relies on the assumption of stochastic Pauli errors, and randomized compilation theory \cite{wallman2015noise} implies that this can be enforced by (1) choosing $\Omega$ so that the distribution of $U(L)$ is invariant under left and right multiplication by Pauli operators, and (2) randomizing $L_{d+1}$ and $L_0$. However, in practice, we find that these conditions on BiRB's circuits are not required, because $\Omega$-distributed circuits rapidly scramble errors. This makes error cancellation negligible after constant depth $k$ \cite{polloreno2023direct}, implying that Eq.~\eqref{eqn:f_d_result} hold to a good approximation for all kinds of small Markovian errors.

\subsection{Deriving the exponential decay model}
\label{sec:fidelity_decay}

Our theory so far shows that $\bar{f}_d$ [Eq.~\eqref{eqn:fbar_d}] is equal to the polarization of depth-$d$ $\Omega$-distributed random circuits multplied by a depth-independent prefactor. Recent work \cite{hines2022demonstrating, polloreno2023direct, proctor2021scalable} has shown that the polarization of $\Omega$-distributed random circuits decays exponentially---from which it follows that $r_{\Omega} \approx \epsilon_{\Omega}$---given the scrambling condition [Eq.~\eqref{eqn:scrambling}] that we require of $\Omega$ and $\mathbb{L}$. This is because Eq.~\eqref{eqn:scrambling} implies that errors within $\Omega$-distributed random circuits cancel with negligible probability, which implies that the polarization of the BiRB core circuit is closely approximated by the product of the polarizations of its constituent layers (the error in this approximation is $O\left(d\varepsilon(\delta+k\varepsilon)\right)$, which is negligible for small $\delta$, where $\delta$ is as defined in Eq.~\eqref{eqn:scrambling} \cite{hines2022demonstrating}). Because the polarizations of $\Omega$-distributed layers approximately multiply, $\bar{f}_d$ decays exponentially, i.e.,
\begin{equation}
    \bar{f}_d \approx Ap^d
\end{equation}
for some $A$ and $p$, and $r_{\Omega} \approx \epsilon_{\Omega}$. 

Here, we give an alternate, complementary proof that $\bar{f}_d$ decays exponentially, which uses the ``$\mathcal{L}$ superchannel'' framework from Refs.~\cite{proctor2017randomized, polloreno2023direct} and is similar to the most accurate theories for standard RB \cite{merkel2018randomized, proctor2017randomized, wallman2017randomized}. We start by expressing $\bar{f}_d$ [Eq.~\eqref{eqn:f_d_simple}] in terms of $d$ applications of a linear operator acting on superoperators (i.e., a ``superchannel''), given by
\begin{equation}
    \mathcal{L}(\mathcal{M}) = \E_{L \in \mathbb{L}} \mathcal{U}(L)^{-1}\mathcal{M}\mathcal{E}_L\mathcal{U}(L). \label{eqn:l_op}
\end{equation}
When $\mathcal{E}_L = \mathbb{I}$ for all $L \in \mathbb{L}$, $\mathcal{L}$ has two unit eigenvalues ($\lambda_0$, $\lambda_1$) and all other eigenvalues ($\lambda_i$, $i > 1$) have absolute value strictly less than 1 \cite{polloreno2023direct}. The following theory requires that the gate errors are sufficiently small that this gap between the unit and non-unit modulus eigenvalues is preserved \footnote{The spectral gap is known to not close when $n$ is small, because Eq.~\eqref{eqn:scrambling} implies that $\Omega$-distributed random circuits rapidly converge to a 2-design. For $n \gg 1$, $\Omega$-distributed random circuits will typically not rapidly converge to a 2-design. Other proofs of exponential decay based on arguing that the polarizations of layer multiply (assuming highly-scrambling layers) do not require this fact, and they apply in the $n \gg 1$ regime.}. For this theory, we do not require that $\mathcal{E}_{d+1}$ and $\mathcal{E}_0$ are global depolarizing channels. Eq.~\eqref{eqn:f_d_simple} can be expressed in terms of $\mathcal{L}$ as
\begin{align}
   \bar{f}_d & = \frac{1}{2^n} \E_{s \in \pauli_n^{\ast}} \Tr\left(s \mathcal{L}^d(\mathcal{E}_{d+1})\left[\mathcal{E}_0[s]\right]\right). \label{eqn:f_l_operator}
\end{align}

In our theory so far, including our definition of $\mathcal{L}$ [Eq.~\eqref{eqn:l_op_channel}], we have used a particular representation of the imperfect gate set---the imperfect gates are given by $\{\mathcal{E}_L\mathcal{U}(L) \mid L \in \mathbb{L} \}$. However, we can express $\bar{f}_d$ in terms of a different representation of these gates with identical predictions, by performing a gauge transformation \cite{Nielsen2020-lt}, i.e., we represent the gates as $\{\mathcal{M}\mathcal{E}_L\mathcal{U}(L)\mathcal{M}^{-1} \mid L \in \mathbb{L}\}$, where $\mathcal{M}$ is an invertible matrix. Below, we re-express the gate set in a particular gauge defined in terms of $\mathcal{L}$. Let $\mathcal{W}=\mathcal{E}_1+\mathcal{E}_{\lambda}$, where $\mathcal{E}_1$ and $\mathcal{E}_{\lambda}$ are eigenoperators of $\mathcal{L}$ with eigenvalues $1$ and $\lambda$, respectively (as defined in Ref.~\cite{polloreno2023direct}, Proposition 3) and where $\lambda$ is the second largest eigenvalue of $\mathcal{L}$. Using the gauge-transformed gate set $\{\mathcal{W}\mathcal{E}_L\mathcal{U}(L)\mathcal{W}^{-1} \mid L \in \mathbb{L}\}$,  Eq.~\eqref{eqn:f_l_operator} becomes
\begin{align}
   \bar{f}_d & = \frac{1}{2^n} \E_{s \in \pauli_n^{\ast}} \Tr\left(s \tilde{\mathcal{L}}^d(\tilde{\mathcal{E}}_{d+1})[\tilde{\mathcal{E}}_0[s]]\right). \label{eqn:f_l_tilde_operator} 
\end{align}
where 
\begin{equation}
    \tilde{\mathcal{L}}[C] = \mathcal{L}[C\mathcal{W}]\mathcal{W}^{-1}, \label{eq:gauge_fixed_L}
\end{equation}
and 
\begin{align}
    \tilde{\mathcal{E}}_{d+1} = \mathcal{E}_{d+1}\mathcal{W}^{-1} \\
    \tilde{\mathcal{E}}_0 = \mathcal{W}\mathcal{E}_0.
\end{align}
If we assume that $\tilde{\mathcal{E}}_{d+1} = \tilde{\mathcal{D}}_{\textrm{meas}}$, where $\tilde{\mathcal{D}}_{\textrm{meas}}$ is a global depolarizing channel (which commutes with all unitary superoperators), it follows from Eq.~\eqref{eqn:f_l_tilde_operator} that
\begin{align}
    \bar{f}_d & = \gamma\left(\tilde{\mathcal{L}}^d[\tilde{\mathcal{D}}_{\textrm{meas}}]\tilde{\mathcal{E}}_0\right) \\
    &= \gamma\left(\tilde{\mathcal{D}}_{\textrm{meas}}\tilde{\mathcal{L}}^d[\mathbb{I}]\tilde{\mathcal{E}}_0\right).
\end{align}
Ref.~\cite{polloreno2023direct} (Proposition 3) shows that $\mathcal{L}(\mathcal{W}) = \mathcal{D}_{\lambda}\mathcal{W}$, where $\mathcal{D}_{\lambda}$ is a global depolarizing channel with polarization $\lambda$. Therefore, $\tilde{\mathcal{L}}^d[\mathbb{I}] = \mathcal{D}_{\lambda^d}$, which implies that 
\begin{align}
    \bar{f}_d & = \gamma(\tilde{\mathcal{D}}_{\textrm{meas}}\mathcal{D}_{\lambda}^d\tilde{\mathcal{E}}_0) \\
    & = \gamma(\tilde{\mathcal{D}}_{\textrm{meas}}\tilde{\mathcal{E}}_0\mathcal{D}_{\lambda}^d) \\
    & = \lambda^d\gamma(\tilde{\mathcal{D}}_{\textrm{meas}}\tilde{\mathcal{E}}_0).
\end{align}
Therefore, $\bar{f}_d$ decays exponentially in depth, at a rate determined by $\lambda$ (the second largest eigenvalue of $\mathcal{L}$). Furthermore, Proposition 4 of Ref.~\cite{polloreno2023direct} implies that $\lambda$ is the average polarization of $\Omega$-distributed layers computed in a particular gauge that is defined by $\mathcal{L}$.

\begin{figure*}[ht!]
    \centering
    \includegraphics{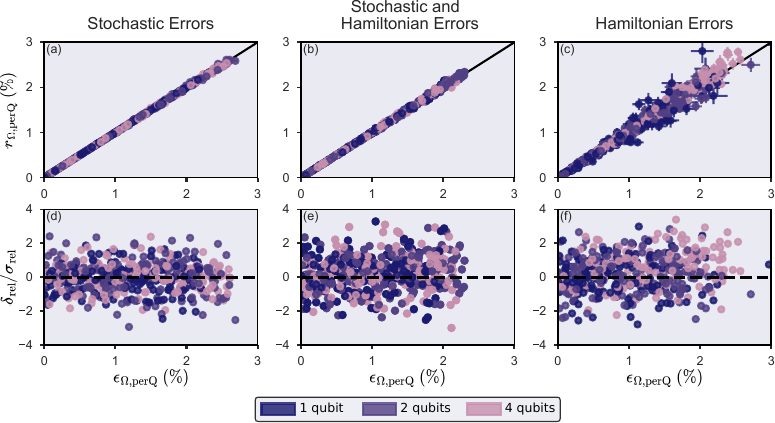}
    \caption{\textbf{Simulations of BiRB for gates with stochastic and Hamiltonian errors.} We simulated BiRB on 1,2, and 4 qubits with randomly-sampled error models. These error models consist of randomly-sampled (a,d) stochastic Pauli errors, (c,f) Hamiltonian errors, (b,e) stochastic and Hamiltonian errors. (a-c): We compare the estimated BiRB error rate $r_{\Omega}$ to $\epsilon_{\Omega}$. Error bars are 1$\sigma$ and are calculated using a standard bootstrap (there are error bars on $\epsilon_{\Omega}$, as well as $r_{\Omega}$, as $\epsilon_{\Omega}$ is estimated via sampling). (d-f): The relative error $\delta_{\textrm{rel}}=\nicefrac{(r_{\Omega}-\epsilon_{\Omega})}{\epsilon_{\Omega}}$, divided by its standard deviation ($\sigma_{\textrm{rel}}$), for each randomly sampled error model. For all error models, we find that $r_{\Omega}$ is approximately equal to $\epsilon_{\Omega}$, and all discrepancies between $r_{\Omega}$ and $\epsilon_{\Omega}$ are consistent with finite sample fluctuations.}
    \label{fig:hs_error_simulations}
\end{figure*}

\section{Simulations}\label{sec:simulations}
In this section, we present simulations of BiRB that show that it reliably estimates the average layer error rate $\epsilon_{\Omega}$. 

\begin{figure*}[t!]
    \centering
    \includegraphics{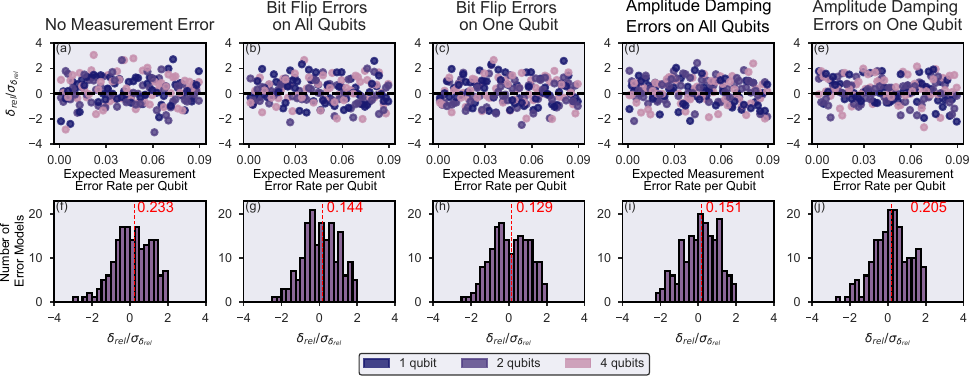}
    \caption{\textbf{Simulations of BiRB with measurement errors.} We simulated BiRB with five types of error models: (a,f) no error on the measurements, (b,g) bit flip errors on the measurements for all $n$ qubits, (c,h) bit flip errors on the measurements for only a single qubit, (d,i) amplitude damping errors on the measurements for all $n$ qubits, and (e,j) amplitude damping errors on the measurement for only a single qubit.  (a-e) The relative error in $r_{\Omega}$ divided by its uncertainty ($\nicefrac{\delta_{\textrm{rel}}}{\sigma_{\delta_{\textrm{rel}}}}$) versus the strength of the measurement error. (f-j) Histograms of $\nicefrac{\delta_{\textrm{rel}}}{\sigma_{\delta_{\textrm{rel}}}}$ for each type of measurement error. We observe no evidence that $r_{\Omega}$ is affected by measurement error, which is consistent with our theory of BiRB and provides further evidence that BiRB is robust to measurement errors.}
    \label{fig:spam}
\end{figure*}

\subsection{BiRB with stochastic and Hamiltonian errors}
\label{sec:hs_sims}
To demonstrate that BiRB accurately estimates $\epsilon_{\Omega}$ under broad conditions, we ran simulations of BiRB with varied error models containing stochastic Pauli and Hamiltionan errors. We simulated BiRB on $n=1,2,$ and $4$ qubits with all-to-all connectivity using the layer set consisting of all possible $n$-qubit layers constructed from parallel applications of $X_{\nicefrac{\pi}{2}}$, $Y_{\nicefrac{\pi}{2}}$, and $\cnot$ gates. These layers were sampled so that the expected density of $\cnot$ gates in a layer is $\xi=\nicefrac{1}{4}$, and each of the two single-qubit gates appears with equal probability. 

We simulated BiRB with three types of error models for these gates: (1) Pauli stochastic errors, (2) Hamiltonian errors, and (3) Pauli stochastic and Hamiltonian errors. To generate each error model, we assign each gate random error rates specified using elementary error generators \cite{blume2021taxonomy}. For each $k$-qubit gate ($k =1, 2$), we specify a post-gate error of the form $e^\mathcal{G}$ for each of $\{X_{\nicefrac{\pi}{2}}, Y_{\nicefrac{\pi}{2}},\cnot\}$, where 
\begin{equation}
    \mathcal{G} = \sum_{i=1}^{4^k-1} s_i \mathcal{S}_i + \sum_{i=1}^{4^k-1} h_i \mathcal{H}_i.
\end{equation}
Here, $\mathcal{S}_1, \mathcal{S}_2, \dots, \mathcal{S}_{4^k-1}$ denote the $k$-qubit stochastic Pauli error generators, and $\mathcal{H}_1, \mathcal{H}_2, \dots, \mathcal{H}_{4^k-1}$ denote the $k$-qubit Hamiltonian error generators. For each error model, we sample $s_i$ and $h_i$ at random (see Appendix ~\ref{app:simulation} for details) to produce a range of expected layer error rates. These models contain no crosstalk errors (but our theory encompasses error models with crosstalk errors) and no state preparation or measurement error.

Figure~\ref{fig:hs_error_simulations} shows the results of these simulations. Fig~\ref{fig:hs_error_simulations}(a)-(c) compares the true average layer error rate per qubit,
\begin{equation}
\epsilon_{\Omega, \,\textrm{perQ}} =  1-(1-\epsilon_{\Omega})^{\nicefrac{1}{n}} \approx \nicefrac{\epsilon_{\Omega}}{n}
\end{equation}
to the estimate of the BiRB error rate per qubit
\begin{equation}
    r_{\Omega, \,\textrm{perQ}} = 1-(1-r_{\Omega})^{\nicefrac{1}{n}} \approx \nicefrac{r_{\Omega}}{n}
\end{equation}
in each simulation, separated into the three families of error models. Error bars (1$\sigma$) are shown, computed using a standard bootstrap (there are error bars on $\epsilon_{\Omega}$ as well as on $r_{\Omega}$ because $\epsilon_{\Omega}$ is computed by random sampling). We observe that for each error model, $r_{\Omega}$ approximately equals $\epsilon_{\Omega}$, as predicted by our theory of BiRB. 

The statistical uncertainty in $r_{\Omega}$ (and $\epsilon_{\Omega}$) is typically much larger in simulations of BiRB experiments on gates with purely Hamiltonian errors, due to higher variance in the performance of circuits of the same depth for this kind of error (as is the case with other RB methods). To quantify any systematic differences between $r_{\Omega}$ and $\epsilon_{\Omega}$, in Fig~\ref{fig:hs_error_simulations}(d)-(f) we show the relative error $\delta_{\textrm{rel}} = (r_{\Omega} - \epsilon_{\Omega})/\epsilon_{\Omega} $ divided by its uncertainty $\sigma_{\textrm{rel}}$, which is computed from $1\sigma$ uncertainties for $r_{\Omega}$ and $\epsilon_{\Omega}$. We see that $r_{\Omega}$ is typically within 2$\sigma$ of $\epsilon_{\Omega}$ for all three classes of error model. The distribution of $\delta_{\textrm{rel}}$ is similar across all error models, suggesting that BiRB is similarly reliable for all three types of error model. Furthermore, we observe that $r_{\Omega}$ does not systematically under- or overestimate $\epsilon_{\Omega}$. This contrasts with the only other method for scalable RB of Clifford gates: MRB. Simulations and theory for MRB both show that MRB systematically underestimates $\epsilon_{\Omega}$ \cite{proctor2021scalable, hines2022demonstrating}. Therefore, our results suggest that BiRB is more accurate than MRB (although note that, unlike BiRB, MRB can scalably benchmark non-Clifford gates).

\subsection{Binary RB with measurement error}
\label{sec:spam_sims}

The simulations presented above (Section~\ref{sec:hs_sims}) did not include SPAM errors, but SPAM errors are often large in current quantum processors. Like other RB protocols, BiRB is designed to be robust to SPAM errors---the effect of SPAM errors is absorbed into a depth-independent prefactor in the exponential fit (see Section~\ref{sec:theory}). Here, we present simulations that demonstrate the robustness of BiRB in the presence of SPAM errors. 

We simulated BiRB on 1, 2, and 4 qubits with single-qubit bit flip and amplitude damping measurement errors. These BiRB simulations used the same layer set and sampling distribution as the simulations presented in Section~\ref{sec:hs_sims}. For these simulations, we simulated BiRB with error models in which the gates have both stochastic Pauli and Hamiltonian errors with rates sampled so that $\epsilon_{\Omega}$ is approximately the same for every error model (see Appendix~\ref{app:simulation} for details). From each set of gate error rates, we construct five error models, each of which has different measurement error. These five error models are: (1) no error on the measurements, (2) bit flip errors on the measurements for all $n$ qubits, (3) bit flip errors on the measurements for only a single qubit, (4) amplitude damping errors on the measurements for all $n$ qubits, and (5) amplitude damping errors on the measurement for only a single qubit. The measurement error rates on each qubit are chosen so that the expected measurement error rate is a constant $p$, which we varied over a range of values (see Appendix~\ref{app:simulation} for details). 

Figure~\ref{fig:spam} shows the results of our simulations of BiRB with measurement errors. We see that the BiRB error rate is not systematically affected by bit flip or amplitude damping error. Figure~\ref{fig:spam} (a-e) shows the relative error ($\delta_{\textrm{rel}}$) in $r_{\Omega}$, divided by its standard deviation ($\sigma_{\delta_{\textrm{rel}}}$), for all error models. We observe no systematic change in $\nicefrac{\delta_{rel}}{\sigma_{\delta_{\textrm{rel}}}}$ as the strength of measurement error ($p$) is varied. Figure~\ref{fig:spam} (f-j) shows the distribution of $\nicefrac{\delta_{rel}}{\sigma_{\delta_{\textrm{rel}}}}$ for all error models with each type of measurement error. We see that the distributions are similar for all types of measurement error. These simulations show no evidence that $r_{\Omega}$ is affected by measurement error, which is consistent with our theory for BiRB. 

\begin{figure*}[t!]
    \centering
    \includegraphics[width=6.5in]{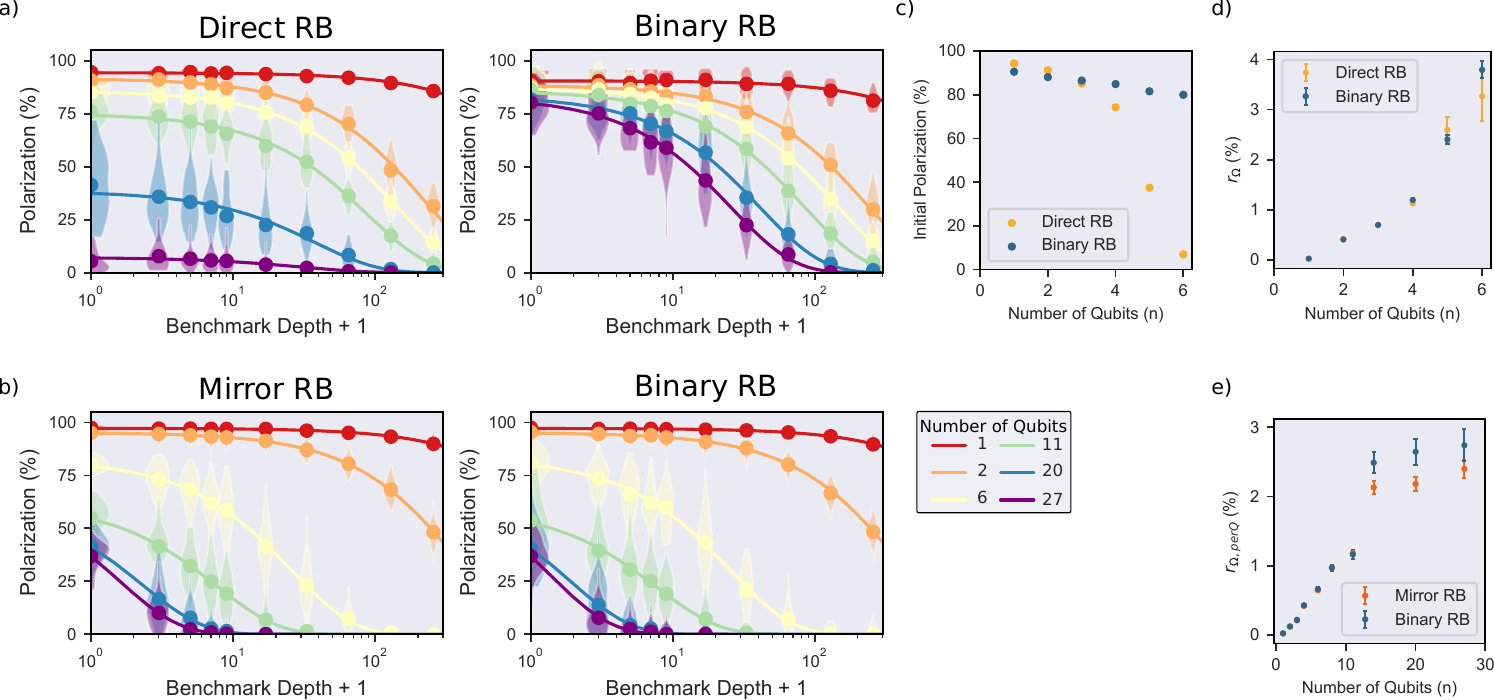}
    \caption{\textbf{BiRB on IBM Q processors.} (a) The results of DRB and BiRB on 1-6 qubits on \texttt{ibm\_perth}. (b) The results of MRB and BiRB on 1-27 qubits on \texttt{ibmq\_kolkata}. (c) DRB's initial (i.e., $d=0$) polarization decreases rapidly as a function of the number of qubits ($n$), due to the random stabilizer state preparation and measurement subroutines in DRB circuits whose size grows quickly with $n$ \cite{proctor2018direct}. In contrast, the initial polarization in BiRB decreases slowly with increasing $n$. (d) The BiRB and DRB error rates ($r_{\Omega}$) are consistent on all qubit subsets. As DRB is a robust technique that is designed to measure the same error rate as BiRB, this provides evidence of BiRB's reliability. (e) The BiRB and MRB error rates are consistent up to $n=11$ qubits, but the BiRB error rate is systematically higher than the MRB error rate for $n > 11$. This is consistent with the theories of BiRB and MRB: MRB theory \cite{proctor2021scalable, hines2022demonstrating} predicts that MRB's $r_{\Omega}$ slightly underestimates $\epsilon_{\Omega}$ 
    (the average layer error rate) whereas BiRB theory predicts that BiRB's $r_{\Omega}$ accurately estimates $\epsilon_{\Omega}$.}
    \label{fig:ibmq_drb}
\end{figure*}

\section{Demonstrations on IBM Q}
\label{sec:demos}

In this section we present demonstrations of BiRB on 7- and 27-superconducting qubit IBM Q devices. We provide experimental evidence that BiRB works by comparing it to two other RB protocols---DRB and MRB---that are designed to measure the same error rate.

\subsection{Validating binary RB in the few-qubit regime}

We ran two experiments comparing BiRB and DRB. We chose to compare the results of BiRB and DRB because (i) DRB is designed to measure the same error rate as BiRB, (ii) BiRB is equivalent to DRB when $n=1$, and (iii) DRB theory \cite{proctor2018direct, polloreno2023direct} shows that DRB is a highly accurate method for estimating the average error rate ($\epsilon_{\Omega}$). 
In these BiRB and DRB experiments, each layer in the core circuit consists of randomly-sampled native $\cnot$ gates (i.e., $\cnot$ gates on connected qubits) and uniformly random single-qubit Clifford gates on all other qubits. We sampled the two-qubit gates using the ``edgegrab'' sampler from Ref.~\cite{proctor2020measuring} with an expected two-qubit gate density of $\xi=\nicefrac{1}{4}$. In our experiment on \texttt{ibm\_perth}, we sampled $K=30$ circuits at exponentially spaced benchmark depths. In our experiment on \texttt{ibm\_hanoi}, we also ran CRB on up to $n=5$ qubits, and we sampled $K=60$ circuits at exponentially spaced benchmark depths. See Appendix~\ref{app:ibmq} for further details.

Fig.~\ref{fig:fig2} and Fig~\ref{fig:ibmq_drb}(a)-(d) show the results of these demonstrations. In all of our BiRB experiments, we observe that the polarization decays exponentially. The DRB and BiRB error rates [Fig.~\ref{fig:fig2}(b) and Fig.~\ref{fig:ibmq_drb}(d)] are consistent with each other on all qubit subsets we tested \footnote{In these DRB experiments, the target circuit outcomes were not randomized, whereas BiRB has randomization of the outcome by design. This is why the distribution of individual circuit polarizations differs between our BiRB and DRB data in Fig.~\ref{fig:ibmq_drb}a, which changes the effect of biased readout error on the distribution of circuit outcomes. However, because this is a depth-independent effect from measurement error, we do not expect the lack of randomization to affect the resulting DRB error rate.} which is consistent with the theories for both DRB and BiRB. These results demonstrate that BiRB is a reliable method for measuring the average layer error rate. In Fig.~\ref{fig:fig2}(b) we also compare the BiRB error rate to an ad hoc heuristic estimate of the average layer error rate obtained by rescaling the results of CRB (see Appendix~\ref{app:ibmq} for details). The rescaled CRB error rate is systematically higher than both the BiRB and DRB error rates. While rescaling CRB error rates to estimate native gate error rates is common practice \cite{mckay2017three, muhonen2015quantifying, barends2014superconducting, raftery2017direct}, this is not typically accurate, as these results demonstrate.

Our results demonstrate that BiRB is more scalable than both DRB and CRB. Although DRB is more scalable than CRB [Fig~\ref{fig:fig2}], the initial (i.e., $d=0$) polarization of DRB circuits [Fig~\ref{fig:ibmq_drb}(c)] still drops off rapidly with increasing $n$, which is due to the $O(\nicefrac{n^2}{\log{n}})$ gate overhead from the stabilizer state preparation and measurement subroutines \cite{polloreno2023direct}. The decrease in initial polarization with $n$ for BiRB circuits is much smaller (note that we expect some decrease in polarization with increased $n$ due to increasing SPAM error).

\subsection{Demonstrating the scalability of binary RB}

To demonstrate that BiRB is reliable in the $n\gg1$ regime, where DRB is infeasible, we ran BiRB and MRB on all 27 qubits of \texttt{ibmq\_kolkata}. MRB is designed to measure the same error rate as BiRB ($\epsilon_{\Omega}$) and it is also scalable. However, the theory of MRB shows that it slightly but systematically underestimates $\epsilon_{\Omega}$ due to correlations between the random layers used in MRB circuits \cite{proctor2021scalable, hines2022demonstrating}. MRB circuits consist of (1) a depth $\nicefrac{d}{2}$ $\Omega$-distributed random circuit, followed by (2) its layer-by-layer inverse, with Pauli frame randomization. MRB theory shows that if the error rates of a $\Omega$-distributed layer and its inverse are uncorrelated, then MRB accurately estimates $\epsilon_{\Omega}$, but that if these error rates are correlated then MRB slightly underestimates $\epsilon_{\Omega}$. In real systems, these error rates are typically correlated.

In the BiRB and MRB circuits we ran, each randomly sampled layer in the core circuit has the form $L = L_1L_2$, where $L_1$ consists of single-qubit gates on all qubits, and $L_2$ consists of parallel $\cnot$ gates on pairs of connected qubits. 
We sampled the single-qubit gates in $L_1$ uniformly from the single-qubit Clifford gates, and we sampled the two-qubit gates in $L_2$ using the ``edgegrab'' sampler \cite{proctor2020measuring} with an expected two-qubit gate density of $\xi=\nicefrac{1}{4}$. We ran circuits with exponentially spaced benchmark depths and sampled $K=60$ circuits of each circuit shape.

Figure~\ref{fig:ibmq_drb}(b) shows the results of our BiRB and MRB experiments on six sets of qubits. Figure~\ref{fig:ibmq_drb}(e) compares the MRB and BiRB error rates for all sets of qubits we tested. The MRB and BiRB error rates are consistent on up to 11 qubits. For $n>11$ qubits, the MRB error rate is systematically lower than the BiRB error rate. This result is consistent with the theory of MRB, which predicts that MRB's $r_{\Omega}$ systematically underestimates $\epsilon_{\Omega}$, and the theory of BiRB, which does not predict a systematic under- or overestimate of $\epsilon_{\Omega}$. MRB theory predicts that MRB's underestimate of $\epsilon_{\Omega}$ is larger when the error rate of a layer and its inverse are highly correlated \cite{proctor2021scalable}. We therefore conjecture that the observed discrepancy between the BiRB and MRB error rates is caused by high variance in the layer error rates in many-qubit circuits, which could occur due to, e.g., large crosstalk error caused by some two-qubit gates. 
The largest difference we observe between MRB and BiRB is in the $n=20$ qubit experiments, where the BiRB error rate is $r_{\Omega} \approx 41.5\%$ and the MRB error rate is $r_{\Omega} \approx 35.7\%$. This discrepancy is consistent with BiRB and MRB theory if the variance in the layers' error rates are sufficiently large. For example, a simple error model that leads to the observed BiRB and MRB error rates is one where each layer experiences purely global depolarizing error and half of the layers have $85\%$ polarization whereas the other half of the layers have $32\%$ polarization.

\section{Discussion}\label{sec:discussion}

In this paper, we introduced BiRB, a highly streamlined RB protocol for Clifford gate sets. Unlike most RB protocols, BiRB does not use motion reversal circuits. Instead, BiRB works by  tracking a single random Pauli operator through each random circuit---using ideas first developed for DFE \cite{Flammia_2011, da_Silva_2011} and later leveraged by Pauli noise learning methods \cite{flammia2021averaged,flammia2019efficient, erhard2019characterizing}. This enables BiRB to scale to many more qubits than most RB methods. Many-qubit BiRB allows for benchmarking of large many-qubit layer sets when individually characterizing all those layers is infeasible, and it is able to accurately capture crosstalk. We have presented a theory for BiRB that proves that BiRB reliably estimates the average error rate of random layers under common assumptions used in RB theory (e.g., Markovian errors), and we have supported this theory with simulations and experimental demonstrations. Our results on IBM Q processors demonstrate BiRB error rates consistent with DRB error rates on up to 6 qubits, and they show that BiRB scales well beyond the limits of DRB and standard CRB.

BiRB enables RB on many more qubits than most existing RB protocols, but it also has advantages in the few-qubit setting. For example, simultaneous few-qubit RB experiments are widely used to quantify crosstalk errors \cite{gambetta2012characterization}, but simultaneous CRB and DRB on $n>1$ qubits are complicated in practice by scheduling problems that arise due to the variable depths of compiled subroutines \cite{mckay2017three}. In contrast, simultaneous BiRB experiments are simple to run because the state preparation and measurement layers in BiRB circuits are each just a single layer of single-qubit gates. Finally, because BiRB does not rely on motion reversal, we anticipate that BiRB can be adapted to benchmark operations that are not intended to be unitary. In particular, in subsequent work we will show that BiRB can be adapted to benchmark gate sets containing mid-circuit measurements---a computational primitive that is essential for quantum error correction.

\section*{Code Availability}
Circuit sampling and data analysis code for BiRB is available in \texttt{pyGSTi} \cite{nielsen2020probing}. Data and code for the simulations and IBM Q demonstrations in this work are available upon reasonable request.

\section*{Acknowledgements} 
This material is based upon work supported by the U.S. Department of Energy, Office of Science, Office of Advanced Scientific Computing Research, Quantum Testbed Pathfinder and the Laboratory Directed Research and Development program at Sandia National Laboratories. This research was funded, in part, by the Office of the Director of National Intelligence (ODNI), Intelligence Advanced Research Projects Activity (IARPA). Sandia National Laboratories is a multi-program laboratory managed and operated by National Technology and Engineering Solutions of Sandia, LLC., a wholly owned subsidiary of Honeywell International, Inc., for the U.S. Department of Energy's National Nuclear Security Administration under contract DE-NA-0003525. We acknowledge the use of IBM Quantum services for this work. All statements of fact, opinion or conclusions contained herein are those of the authors and should not be construed as representing the official views or policies of the U.S. Department of Energy, or the U.S. Government, or IBM, or the IBM Quantum team. This written work is authored by an employee of NTESS. The employee, not NTESS, owns the right, title and interest in and to the written work and is responsible for its contents. Any subjective views or opinions that might be expressed in the written work do not necessarily represent the views of the U.S. Government. The publisher acknowledges that the U.S. Government retains a non-exclusive, paid-up, irrevocable, world-wide license to publish or reproduce the published form of this written work or allow others to do so, for U.S. Government purposes. The DOE will provide public access to results of federally sponsored research in accordance with the DOE Public Access Plan.

\bibliography{Bibliography}

\appendix
\onecolumngrid

\section{Clifford group binary RB}\label{app:standard_rb}
In this appendix, we discuss Clifford group BiRB---i.e., BiRB with uniformly random $n$-qubit Clifford layers. Furthermore, we use 2-design twirling theory to show that that the expected polarization $(\bar{f}_d)$ decays exponentially in depth for Clifford group BiRB.

\subsection{Clifford group binary RB protocol}
We start by introducing the Clifford group BiRB protocol. A depth-$d$ Clifford group BiRB circuit is a circuit $ C= C_dC_{d-1}\cdots C_1C_0$ of $d+1$ random $n$-qubit Clifford gates $C_i \in \mathbb{C}_n$. The first random Clifford $C_0$ produces a uniformly random $n$-qubit stabilizer state $\ket{\psi}^{\otimes n} = C_0\ket{0}^{\otimes n}$. Picking a uniformly random non-Identity element of the stabilizer group of $\ket{\psi}$ is equivalent to picking a uniformly random element of $\mathbb{P}_n^{\ast}$, which allows us to do Clifford group BiRB without the initial tensor product state preparation. At the end of the circuit, the evolved Pauli operator is
\begin{equation}
    s'_C=U(C_d\cdots C_2C_1)sU(C_1^{-1}\cdots C_{d-1}^{-1}C_d^{-1}). \label{eqn:s_prime}
\end{equation} 
For simplicity, we will assume the ability to measure in all Pauli bases in the following discussion, so that we do not need a final layer of gates to transform $s'$ into a tensor product of $Z$ and $I$ Pauli operators. We note that the method stated here is equivalent to BiRB as stated in the main text, but with the first layer of gates $L_0$ recompiled into the first benchmarking layer (here, $C_0$). We choose to discuss this variant because it uses circuits consisting entirely of i.i.d.~layers.

Our protocol is the following:
\begin{enumerate}
    \item For a range of integers $d \geq 0$, sample $K$ circuits $C_dC_{d-1}\cdots C_1C_0$. For each circuit, sample a random  $s \neq \Id_n$ in the stabilizer group of $U(C_0)\ket{0}^{\otimes n}$. 
    \item Run each circuit $C$ $N\geq 1$ times and compute $\langle s'_C \rangle $. Then, compute the average over all circuits of benchmark depth $d$, 
    \begin{equation}
        \bar{f}_d = \frac{1}{K}\sum\limits_{C_d} \langle s'_C \rangle. \label{app:fbar_d}
    \end{equation}
    \item Fit $\bar{f}_d$ to an exponential, $\bar{f}_d = Ap^d$, where $A$ and $p$ are fit parameters. The RB error rate (scaled to correspond to average gate infidelity) is given by 
    \begin{equation}
   r = (2^n - 1)(1 - p)/2^n.
   \end{equation}
\end{enumerate}

\subsection{Extracting the RB error rate in Clifford group binary RB}
We now show that $\bar{f}_d$ decays exponentially in benchmark depth using 2-design twirling. We assume arbitrary gate-independent Markovian error on each $n$-qubit Clifford---i.e., $\phi(C_i) = \mathcal{E}\mathcal{U}(C_i)$ for all $C_i$. An imperfect implementation of a benchmark depth-$d$ Clifford group BiRB circuit is given by
\begin{align}
\phi(C) & = \phi(C_d)\cdots\phi(C_2)\phi(C_1)\phi(C_0) \\
& =  \mathcal{E} \mathcal{U}(C_d) \mathcal{E} \mathcal{U}(C_{d-1}) \cdots \mathcal{U}(C_1) \mathcal{E} \mathcal{U}(C_0). \label{eqn:standard_rb_circuit}
\end{align}
Here, we will use $S_{C_0}$ to denote the stabilizer group of $U(C_0)\ket{0}^{\otimes n}$. The expected polarization of a benchmark depth $d$ circuit is
\begin{align}
    \bar{f}_d & = \E_{C_0,\dots, C_{d}} \E_{s \in S^{\ast}_{C_0}} \Tr\left(s'_C\phi(C)[(\ket{0}\bra{0})^{\otimes n}]\right), \label{eqn:f_rb_0} \\
    & = \E_{C_0,\dots, C_{d}} \E_{s \in S^{\ast}_{C_0}} \Tr\left(\mathcal{U}(C_d\cdots C_1)[s]\phi(C)[(\ket{0}\bra{0})^{\otimes n}]\right)  \label{eqn:f_rb_1}\\
    & = \E_{C_0,\dots, C_{d}} \E_{s \in S^{\ast}_{C_0}} \Tr\left(s\mathcal{U}(C_1^{-1}\cdots C_d^{-1})[\phi(C)[(\ket{0}\bra{0})^{\otimes n}]]\right).  \label{eqn:f_rb_2}
\end{align}
Eq.~\eqref{eqn:f_rb_1} follows from applying the definition of $s'_C$ [Eq.~\eqref{eqn:s_prime}] to Eq.~\eqref{eqn:f_rb_0}, and Eq.~\eqref{eqn:f_rb_2} follows from Eq.~\eqref{eqn:f_rb_1} and the cyclic property of the trace. 
Furthermore, we have
\begin{equation}
    \mathcal{U}(C_1^{-1}\cdots C_d^{-1})\phi(C)  = \mathcal{U}(C_1^{-1}) \cdots \mathcal{U}(C_d^{-1}) \mathcal{E} \mathcal{U}(C_d) \mathcal{E} \mathcal{U}(C_{d-1}) \cdots \mathcal{U}(C_1) \mathcal{E} \mathcal{U}(C_0).
\end{equation}
Therefore, averaging over $C_1, \cdots C_d$ in Eq.~\eqref{eqn:f_rb_2} twirls the error channels $\mathcal{E}$ into global depolarizing error \cite{dankert2009exact}. Eq.~\eqref{eqn:f_rb_2} becomes
\begin{equation}
    \bar{f}_d  = \E_{C_{0}} \E_{s \in S^{\ast}_{C_0}} \Tr\left(s \tilde{\mathcal{E}}^d [\phi(C_0)[(\ket{0}\bra{0})^{\otimes n}]]\right), \label{eqn:f_rb_3}
\end{equation}
where $\tilde{\mathcal{E}} = \E_{C \in \mathbb{C}_n} \mathcal{U}(C)^{-1}\mathcal{E}\mathcal{U}(C)$. Since $\mathcal{E}$ is perfectly twirled by the $n$-qubit Clifford group, $\tilde{\mathcal{E}}$ is an $n$-qubit depolarizing error channel $\tilde{\mathcal{E}}[\rho] =\gamma\rho + (1-\gamma) \nicefrac{\Id_n}{2^n}$. Using this result in Eq.~\eqref{eqn:f_rb_3}, we find that
\begin{align}
    \bar{f}_d & = \gamma^d  \E_{C_{0}} \E_{s \in S^{\ast}_{C_0}}  \Tr\left(s \phi(C_0) [(\ket{0}\bra{0})^{\otimes n}]\right)  \nonumber \\
    & + \frac{1}{2^n} (1-\gamma^d)  \E_{C_{0}} \E_{s \in S^{\ast}_{C_0}}  \Tr\left(s\phi(C_0)[\Id_n]\right) \\
    & = A\gamma^d,
\end{align}
where 
\begin{equation}
    A = \E_{C_{0}} \E_{s \in S^{\ast}_{C_0}}  \Tr\left(s \phi(C_0) [(\ket{0}\bra{0})^{\otimes n}]\right).
\end{equation}

Therefore, $\bar{f}_d$ decays exponentially in circuit depth, at a rate determined by the fidelity of $\mathcal{E}$. This implies that the Clifford group BiRB error rate is the same as the (standard) CRB error rate. 

\section{Binary RB statistics}
\label{app:statistics}
In this appendix, we show that the number of circuits required for the BiRB protocol is independent of the number of qubits $n$. We work in the single-shot limit, so that each measurement result is an independent random variable $f_i \in [-1, 1]$. At each circuit depth, we compute the estimate $\hat{f}_d = \frac{1}{K}\sum_{i=1}^K f_i$ of the expected polarization of benchmark depth-$d$ BiRB circuits ($\bar{f}_d$). Hoeffding's inequality says that 
\begin{align}
    Pr\left[\mid\hat{f}_d-\bar{f}_d\mid \geq \delta\right] \leq 2\exp\left(-\frac{1}{2}\delta^2 K\right),
\end{align}
where $K$ is the number of circuits ran. We have $\bar{f}_d \approx A\overline{\gamma}^d$, where $\overline{\gamma} = \E_L \gamma(\mathcal{E}_L)$ is the expected layer polarization. Replacing $\delta$ with a relative uncertainty $\delta = \alpha A\overline{\gamma}^d$, we have
\begin{align}
    Pr\left[\mid\hat{f}_d-\bar{f}_d\mid \geq \alpha\overline{\gamma}^d\right] \leq 2\exp\left(-\frac{1}{2}\alpha^2A^2\overline{\gamma}^{2d} K\right).
\end{align}
The number of circuits required to obtain an estimate of $\bar{f}_d$ to within relative uncertainty uncertainty $\alpha$ with probability at least $1-\nu$ is therefore 
\begin{equation}
    K = \frac{2\log{(2/\nu)}}{\alpha^2A^2\overline{\gamma}^{2d}}.
\end{equation}
Importantly, this does not scale with the number of qubits $n$.

In order to obtain an accurate estimate of the decay rate of $\bar{f}_d$, we need to estimate $\bar{f}_d$ for at least two depths $d_0$, $d_1$ with $d_1-d_0 = O(\log(\nicefrac{1}{\bar{\gamma}}))$. For simplicity, we take $d_0=0$ and $d_1=\log(\nicefrac{1}{\bar{\gamma}})$. We consider the simplified scenario of estimating $\bar{\gamma}$ using the ratio of these two polarization estimates,
\begin{equation}
    \bar{\gamma} = \left(\frac{f_{d_1}}{f_{d_0}}\right)^{\frac{1}{d_1}}.
\end{equation}
In order to estimate $\bar{\gamma}$ to multiplicative accuracy $\beta$, we need to estimate $f_{d_1}$ and $f_{d_0}$ to multiplicative accuracy $\nicefrac{d_1 \beta}{2}$. At depth $d_1$, the number of shots required is 
\begin{equation}
    K(d_1) =  \frac{8\log{(2/\nu)}}{d_1^2\beta^2A^2\overline{\gamma}^{2d_1}},
\end{equation}
and the number of shots required at depth $0$ is 
\begin{equation}
    K(d_0) =  \frac{8\log{(2/\nu)}}{d_1^2\beta^2A^2}.
\end{equation}

\section{Stabilizers of tensor product states}
\label{app:theory}
Here, we prove the result used in Section~\ref{sec:l_op} to go from Eq.~\eqref{eqn:f_stabilizer_expansion} to Eq.~\eqref{eqn:f_d_simple}: For any $s,p \in \mathbb{P}_n^{\ast}$ with $s \neq \pm p$ and $[s,p]=0$, there is a bijection between tensor product stabilizer states $\ket{\psi(s)}$ that are stabilized by $p$ and tensor product stabilizer states  $\ket{\psi(s)}$ that are stabilized by $-p$.

To construct our bijection, we pick an ordering of the $n$ qubits, and we express $s$ and $p$ as tensor products of single qubit Pauli operators---i.e., $s = \otimes_{i=0}^n s_i$ and $p = \otimes_{i=0}^n p_i$. For any $\ket{\psi(s)}=L_0\ket{0}^{\otimes n}$ satisfying $s'\ket{\psi(s)}=\ket{\psi(s)}$, we can create another tensor product state $\ket{\psi'(s)}=L_0'\ket{0}^{\otimes n}$ satisfying $-p'\ket{\psi'(s)}=\ket{\psi'(s)}$ as follows: We start with $L_0'=L_0$ and modify some of the single-qubit gates in $L_0'$. Find the lowest index $j$ such that $s_j = I$ and $p_j \neq I$, if it exists. There are two cases:
\begin{enumerate}
\item Such a $j$ exists: Because $p_j \neq I$, the gate on qubit $j$ produces a $+1$ eigenstate of some $q \in \pauli_n^{\ast}$. Replace this gate in $L_0'$ with a gate that produces a $+1$ eigenstate of $-q$.
\item No such $j$ exists: If there is no $j$ such that $s_j = I$ and $p_j \neq I$, then there must be a qubit $j$ such that $s_j \neq I$ and $p_j = I$. The gate on qubit $j$ produces a $+1$ eigenstate of some $q \in \pauli_n^{\ast}$. Replace this gate in $L_0'$ with a gate that produces a $+1$ eigenstate of $-q$. In addition, because we know that $p \neq I$, there must be some other qubit $j'$ such that $p_{j'} \neq I$, and it must be that $s_{j'}=p_{j'}$. Suppose the gate on qubit $j'$ produces a $+1$ eigenstate of some $q' \in \pauli_n^{\ast}$. Replace this gate with a gate that produces a $+1$ eigenstate of $-q'$.
\end{enumerate}
The new state $\ket{\psi'(s)}$ produced by the new layer of gate is stabilized by $s$ and $-p$, and this mapping is bijective. From this result it follows that
\begin{equation}
    \sum_{\substack{s' \in S_{\psi}\\ s' \neq I, s}} s' = 0, 
\end{equation}
from which Eq.~\eqref{eqn:f_d_simple} follows. 

\section{Simulations of binary RB} 
\label{app:simulation}
\subsection{Binary RB with stochastic Pauli and Hamiltonian errors}
\label{app:hs_sims}

We simulated BiRB on $n=1,2,4$ qubits with all-to-all connectivity using layers constructed from the gate set $\{X_{\nicefrac{\pi}{2}}, Y_{\nicefrac{\pi}{2}}, \cnot\}$. The error models we use in our BiRB simulations are defined in terms of the stochastic and Hamiltonian elementary error generators defined in Ref.~\cite{blume2021taxonomy}. For each $k$-qubit gate ($k =1, 2$), we specify a post-gate error of the form $e^\mathcal{G}$ for each of $\{X_{\nicefrac{\pi}{2}}, Y_{\nicefrac{\pi}{2}},\cnot\}$, where 
\begin{equation}
    \mathcal{G} = \sum_{i=1}^{4^k-1} s_i \mathcal{S}_i + \sum_{i=1}^{4^k-1} h_i \mathcal{H}_i,
\end{equation}
where $\mathcal{S}_1, \mathcal{S}_2, \dots, \mathcal{S}_{4^k-1}$ denote the $k$-qubit stochastic Pauli error generators, and $\mathcal{H}_1, \mathcal{H}_2, \dots, \mathcal{H}_{4^k-1}$ denote the $k$-qubit stochastic Pauli error generators. For each error model, we sample $s_i$ and $h_i$ at random to produce a range of expected layer error rates. To generate error models, we start with an overall error parameter $p$ that determines the expected gate error rates in the model. We generate models with $p \in [0, 0.01875]$ for 150 evenly-spaced values for the single-qubit models and $p \in [0, 0.0750]$ for 150 evenly-spaced values for the 2- and 4-qubit models. We use $p$ to determine the expected rates of stochastic and Hamiltonian errors. In the stochastic Pauli error models, we set $h=0$ and $s=1.2p$. In the Hamiltonian error models, we set $s=0$ and $h=\sqrt{8p}$ for $n=4$ qubit models, and we set $s=0$ and $=\sqrt{6p}$ for $n=1,2$ qubit models. In the stochastic Pauli and Hamiltonian error models, we generate $s \in [0, p]$ at random and set $h = \sqrt{2p-s}$. These sampling parameters are chosen to produce models with a similar range of per-qubit error rates across all error model types and all values of $n$.

We include qubit-dependent Hamiltonian errors and stochastic Pauli errors on each gate, with Hamiltonian error rates sampled in the range $[0, \chi h]$, and stochastic Pauli error rates sampled in the range $[0, \chi s]$, where $\chi=0.1$ if $n=2,4$ and $k=1$ (i.e., single-qubit gate error rates are sampled so that their expected error rate is $\nicefrac{1}{10}$ the error rate of 2-qubit gates) and $\chi=1$ otherwise. The stochastic and Hamiltonian errors are each split randomly across the $4^k-1$ error generators.

For each error model, we run $K = 100$ BiRB circuits at each depth $d \in \{0\} \cup \{2^j \mid 0 \leq j \leq 8\}$. Each layer in the core circuit consists of randomly-sampled CNOT gates and uniformly random gates from the set $\{X_{\nicefrac{\pi}{2}}, Y_{\nicefrac{\pi}{2}}, I\}$ on all other qubits. We sampled the two-qubit gates using the ``edgegrab'' sampler from Ref.~\cite{proctor2020measuring} with an expected two-qubit gate density of $\xi=\nicefrac{1}{2}$. 

We also approximate the average layer error rate $\epsilon_{\Omega}$ via sampling. We sample $K = 100$ $\Omega$-distributed random circuits at each depth $d \in \{0\} \cup \{2^j \mid 0 \leq j \leq 8\}$ (using the same layer sampling as described above) and determine their polarization, then fit the resulting data to an exponential to obtain an estimate of $\epsilon_{\Omega}$.

\subsection{Binary RB with measurement errors}

We simulated BiRB on $n=1, 2, 4$ qubits with single-qubit bit flip and amplitude damping measurement error. These BiRB circuits used layers constructed from the gates $\{X_{\nicefrac{\pi}{2}}, Y_{\nicefrac{\pi}{2}}, \cnot\}$. In these simulations, we simulated BiRB with error models in which the gates have both stochastic Pauli and Hamiltonian errors. We generated $30$ models with Hamiltonian and stochastic errors. Each error model had randomly-chosen error rates sampled so that the expected stochastic error rate was $\nicefrac{p}{2}$ and the expected Hamiltonian error rate was $\sqrt{\nicefrac{p}{2}}$, and we set $p=0.015n$. In our $n=2,4$ qubit simulations, we sampled the errors on single-qubit gates so that their expected error rates were approximately $\nicefrac{1}{10}$ of the expected two-qubit gate error rate.

From each set of gate error rates, we construct five error models, each of which has different measurement error. These five error models are: (1) no error on the measurements, (2) bit flip errors on the measurements for all $n$ qubits, (3) bit flip errors on the measurements for only a single qubit, (4) amplitude damping errors on the measurements for all $n$ qubits, and (5) amplitude damping errors on the measurement for only a single qubit. We define our measurement error using the  single-qubit elementary error generators $S_X$, $S_Y$, and $A_{X,Y}$ defined in Ref.~\cite{blume2021taxonomy}, and an error strength parameter $p_m$. In our bit flip error models, we add the error $\mathcal{E} = e^{p_m S_x}$ immediately before the measurement.  In our amplitude damping error models, we add the error  $\mathcal{E} = e^{p_m(S_X+S_Y+A_{X,Y})}$ immediately before measurement. In error models with measurement error on a single qubit, we generate error models with $60$ evenly-spaced values of $p_m \in [0.0001,0.09]$. In error models with measurement error on all qubits we sample a uniform random $p_m \in [0,\nicefrac{2p}{n}]$ independently for each qubit, for $60$ evenly-spaced values of $p \in [0.0001,0.09]$,

For each error model, we run $K = 100$ BiRB circuits at each depth $d \in \{0\} \cup \{2^j \mid 1 \leq j \leq 8\}$ using the same gate set and layer sampling distribution as in Appendix~\ref{app:hs_sims}. We approximate $\epsilon_{\Omega}$ via sampling using the method described in Appendix~\ref{app:hs_sims}.

\section{Demonstrating binary RB on IBM Q}
\label{app:ibmq}
\subsection{Details of demonstration on IBM Hanoi}
We ran BiRB and other RB protocols on \texttt{ibm\_hanoi}, \texttt{ibm\_perth}, and \texttt{ibmq\_kolkata}. In this appendix, we provide the details of our RB experiments on \texttt{ibm\_hanoi}. Details of our experiments on \texttt{ibm\_perth} and \texttt{ibmq\_kolkata} can be found in Section~\ref{sec:demos}.

We ran DRB, BiRB, and CRB on \texttt{ibm\_hanoi} [Fig.~\ref{fig:fig2}]. For our DRB and BiRB experiments, we benchmarked a gate set consisting of the 24 single-qubit Clifford gates and CNOT. Each benchmarking layer consisted of random CNOT gates, respecting the device connectivity, and random single-qubit gates on all other qubits. The CNOT gates were sampled using edgegrab sampling with expected two-qubit gate density of $\xi = \nicefrac{1}{4}$. We ran $K=60$ circuits at exponentially spaced benchmark depths for each of DRB, BiRB, and CRB. We randomized the order of the circuit list and ran each circuit with 1000 shots. 

The CRB error rate is an estimate of the average error rate of a (compiled) $n$-qubit Clifford gate. To directly compare the CRB error rate to the DRB and BiRB error rates, we use a heuristic to approximate the average error of a layer from the distribution $\Omega$ we used to sample the DRB and BiRB circuits. Our estimate for the $n$-qubit layer error rate is $r_{\Omega, \textrm{est}} = (r_n/k_n)n\xi$, where $r_n$ is the $n$-qubit CRB error rate and $k_n$ is the average number of two-qubit gates per $n$-qubit Clifford. 

\subsection{RB error rates and calibration data}

Here, we provide the RB error rates and device calibration data from all of our RB experiments on \texttt{ibm\_perth} [Tables \ref{tab:ibm_perth_r} and \ref{tab:ibm_perth_calibration} ], \texttt{ibmq\_kolkata} [Table \ref{tab:ibmq_kolkata_r} and \ref{tab:ibmq_kolkata_calibration}], \texttt{ibm\_hanoi} [Table \ref{tab:ibm_hanoi_r} and \ref{tab:ibm_hanoi_calibration}].

\begin{table*}[h!]
\centering
\begin{tabular}{|p{6cm}|l|l|}
\cline{1-3}
qubit subset & $r_{\Omega}$ (BiRB)&  $r_{\Omega}$ (DRB) \\\cline{1-3}
\Q0 &  0.027(1)& 0.0279(5)  \\\cline{1-3} 
(\Q0, \Q1) &  0.41(1) & 0.42(1)  \\\cline{1-3} 
(\Q0, \Q1, \Q2) & 0.70(2)  & 0.70(1)  \\\cline{1-3} 
(\Q0, \Q1, \Q2, \Q3) & 1.20(4) & 1.13(3)\\\cline{1-3} 
(\Q0, \Q1, \Q2, \Q3, \Q5) & 2.40(9) & 2.6(2) \\\cline{1-3} 
(\Q0, \Q1, \Q2, \Q3, \Q5, \Q6) &  3.8(2)& 3.3(4) \\\cline{1-3} 
(\Q0, \Q1, \Q2, \Q3, \Q4, \Q5, \Q6) & 5.6(3) & fit failed \\\cline{1-3} 
\end{tabular}
    \caption{\textbf{BiRB and DRB on IBM Perth} The RB error rates from every BiRB and DRB experiment we ran on \texttt{ibm\_perth}.}
    \label{tab:ibm_perth_r}
\end{table*}

\begin{table*}[h!]
\begin{tabular}{|l|l|l|l|l|l|l|l|l|}
\cline{1-9}
qubit & $T_1$ (us) & $T_2$ (us) & frequency (GHz)& anharmonicity  (GHz) & readout error & Pr(prep 1, measure 0)& Pr(prep 0, measure 1)& readout length  (ns) \\\cline{1-9}
\Q0 & 122.58 & 84.49 & 5.16 & -0.34 & 0.019 & 0.018 & 0.020 & 675.56 \\\cline{1-9}
\Q1 & 96.44 & 36.20 & 5.03 & -0.34 & 0.019 & 0.022 & 0.016 & 675.56 \\\cline{1-9}
\Q2 & 298.92 & 61.66 & 4.86 & -0.35 & 0.010 & 0.011 & 0.009 & 675.56 \\\cline{1-9}
\Q3 & 170.68 & 179.99 & 5.13 & -0.34 & 0.012 & 0.016 & 0.008 & 675.56 \\\cline{1-9}
\Q4 & 77.10 & 109.16 & 5.16 & -0.33 & 0.012 & 0.011 & 0.012 & 675.56 \\\cline{1-9}
\Q5 & 148.18 & 69.49 & 4.98 & -0.35 & 0.014 & 0.015 & 0.013 & 675.56 \\\cline{1-9}
\Q6 & 167.33 & 245.76 & 5.16 & -0.34 & 0.007 & 0.008 & 0.006 & 675.56 \\\cline{1-9}
\end{tabular}
\caption{\textbf{IBM Perth calibration data.} Calibration data from \texttt{ibm\_perth} from the time of our BiRB demonstrations.}
    \label{tab:ibm_perth_calibration}
\end{table*}

\begin{table*}[h!]
\centering
\begin{tabular}{|p{6cm}|l|l|}
\cline{1-3}
qubit subset & $r_{\Omega}$ (BiRB)&  $r_{\Omega}$ (MRB) \\\cline{1-3}
\Q0 &  0.0230(4) & 0.0228(3)\\\cline{1-3}
(\Q0, \Q1) & 0.245(3)& 0.246(3) \\\cline{1-3}
(\Q0, \Q1, \Q2) & 0.65(1) & 0.63(1)  \\\cline{1-3} 
(\Q0, \Q1, \Q2, \Q3) & 1.71(4)  & 1.67(4) \\\cline{1-3} 
(\Q0, \Q1, \Q2, \Q3, \Q4, \Q5) &  3.9(1)& 3.8(1) \\\cline{1-3} 
(\Q0, \Q1, \Q2, \Q3, \Q4, \Q5, \Q6, \Q7) &  7.5(3)& 7.5(3) \\\cline{1-3} 
(\Q0, \Q1, \Q2, \Q3, \Q4, \Q5, \Q6, \Q7, \Q8, \Q9, \Q10) & 12.1(6) & 12.2(4) \\\cline{1-3} 
(\Q0, \Q1, \Q2, \Q3, \Q4, \Q5, \Q6, \Q7, \Q8, \Q9, \Q10, \Q11, \Q12, \Q13) & 30(2) & 26.0(9) \\\cline{1-3} 
(\Q0, \Q1, \Q2, \Q3, \Q4, \Q5, \Q6, \Q7, \Q8, \Q9, \Q10, \Q11, \Q12, \Q13, \Q14, \Q15, \Q16, \Q17, \Q18, \Q19) & 42(2) & 36(1)  \\\cline{1-3} 
(\Q0, \Q1, \Q2, \Q3, \Q4, \Q5, \Q6, \Q7, \Q8, \Q9, \Q10, \Q11, \Q12, \Q13, \Q14, \Q15, \Q16, \Q17, \Q18, \Q19, \Q20, \Q21, \Q22, \Q23, \Q24, \Q25, \Q26) & 53(3) & 48(2) \\\cline{1-3} 
\end{tabular}
    \caption{\textbf{BiRB and MRB on IBMQ Kolkata.} The RB error rates for every BiRB and MRB experiment we ran on \texttt{ibmq\_kolkata}.}
    \label{tab:ibmq_kolkata_r}
\end{table*}

\begin{table*}[h!]
\begin{tabular}{|l|l|l|l|l|l|l|l|l|}
\cline{1-9}
qubit & $T_1$ (us) & $T_2$ (us) & frequency (GHz)& anharmonicity  (GHz) & readout error & Pr(prep 1, measure 0)& Pr(prep 0, measure 1)& readout length  (ns) \\\cline{1-9}
\Q0 & 126.85 & 24.35 & 5.20 & -0.34 & 0.011 & 0.010 & 0.012 & 675.56 \\\cline{1-9}
\Q1 & 152.85 & 147.77 & 4.99 & -0.35 & 0.014 & 0.013 & 0.014 & 675.56 \\\cline{1-9}
\Q2 & 87.82 & 37.07 & 5.11 & -0.34 & 0.012 & 0.016 & 0.008 & 675.56 \\\cline{1-9}
\Q3 & 113.23 & 47.66 & 4.87 & -0.35 & 0.043 & 0.061 & 0.026 & 675.56 \\\cline{1-9}
\Q4 & 120.73 & 104.49 & 5.22 & -0.34 & 0.031 & 0.032 & 0.031 & 675.56 \\\cline{1-9}
\Q5 & 120.73 & 42.50 & 5.10 & -0.34 & 0.027 & 0.026 & 0.027 & 675.56 \\\cline{1-9}
\Q6 & 118.32 & 65.44 & 5.20 & -0.34 & 0.031 & 0.028 & 0.034 & 675.56 \\\cline{1-9}
\Q7 & 128.19 & 23.16 & 5.02 & -0.35 & 0.065 & 0.026 & 0.104 & 675.56 \\\cline{1-9}
\Q8 & 171.87 & 193.75 & 4.96 & -0.35 & 0.017 & 0.022 & 0.013 & 675.56 \\\cline{1-9}
\Q9 & 168.79 & 57.68 & 5.06 & -0.34 & 0.155 & 0.165 & 0.144 & 675.56 \\\cline{1-9}
\Q10 & 141.42 & 67.37 & 5.18 & -0.34 & 0.017 & 0.017 & 0.018 & 675.56 \\\cline{1-9}
\Q11 & 11.32 & 17.23 & 4.88 & -0.37 & 0.042 & 0.052 & 0.031 & 675.56 \\\cline{1-9}
\Q12 & 140.45 & 54.01 & 4.96 & -0.35 & 0.015 & 0.025 & 0.006 & 675.56 \\\cline{1-9}
\Q13 & 130.20 & 154.80 & 5.02 & -0.35 & 0.011 & 0.012 & 0.011 & 675.56 \\\cline{1-9}
\Q14 & 165.62 & 120.93 & 5.12 & -0.34 & 0.007 & 0.009 & 0.004 & 675.56 \\\cline{1-9}
\Q15 & 150.20 & 162.20 & 5.03 & -0.34 & 0.008 & 0.009 & 0.006 & 675.56 \\\cline{1-9}
\Q16 & 88.63 & 63.20 & 5.22 & -0.34 & 0.017 & 0.014 & 0.019 & 675.56 \\\cline{1-9}
\Q17 & 100.91 & 33.29 & 5.24 & -0.34 & 0.006 & 0.007 & 0.004 & 675.56 \\\cline{1-9}
\Q18 & 120.14 & 57.40 & 5.09 & -0.34 & 0.011 & 0.008 & 0.014 & 675.56 \\\cline{1-9}
\Q19 & 132.28 & 117.35 & 5.00 & -0.34 & 0.011 & 0.013 & 0.009 & 675.56 \\\cline{1-9}
\Q20 & 135.13 & 155.12 & 5.19 & -0.34 & 0.008 & 0.010 & 0.007 & 675.56 \\\cline{1-9}
\Q21 & 115.65 & 103.94 & 5.27 & -0.34 & 0.005 & 0.005 & 0.005 & 675.56 \\\cline{1-9}
\Q22 & 149.22 & 42.50 & 5.12 & -0.34 & 0.010 & 0.013 & 0.008 & 675.56 \\\cline{1-9}
\Q23 & 153.04 & 129.84 & 5.14 & -0.34 & 0.006 & 0.007 & 0.006 & 675.56 \\\cline{1-9}
\Q24 & 136.85 & 30.46 & 5.00 & -0.35 & 0.011 & 0.017 & 0.005 & 675.56 \\\cline{1-9}
\Q25 & 266.91 & 163.80 & 4.92 & -0.35 & 0.007 & 0.011 & 0.004 & 675.56 \\\cline{1-9}
\Q26 & 138.55 & 100.89 & 5.12 & -0.34 & 0.007 & 0.010 & 0.003 & 675.56 \\\cline{1-9}
\end{tabular}
\caption{\textbf{IBMQ Kolkata calibration data.} Calibration data from \texttt{ibmq\_kolkata} from the time of our BiRB demonstrations.}
    \label{tab:ibmq_kolkata_calibration}
\end{table*}

\begin{table*}[h!]
\centering
\begin{tabular}{|p{6cm}|l|l|l|}
\cline{1-4}
qubit subset & $r_{\Omega}$ (BiRB)&  $r_{\Omega}$ (DRB) & $r$ (CRB) \\\cline{1-4}
Q0 & 0.0224(3) & 0.0222(3)  & 0.0318(3) \\\cline{1-4} 
(Q0, Q1) & 0.296(5)& 0.300(5)  & 2.13(3) \\\cline{1-4} 
(Q0, Q1, Q2) & 0.456(7) & 0.445(8)  & 10.9(2)\\\cline{1-4} 
(Q0, Q1, Q2, Q3) & 0.94(2)& 0.98(3)  & 55(2)\\\cline{1-4} 
(Q0, Q1, Q2, Q3, Q4) & 1.61(3) & 1.53(4) & 99(1)\\\cline{1-4} 
(Q0, Q1, Q2, Q3, Q4, Q5) & 1.82(4) & 1.9(2)  &\\\cline{1-4} 
('Q0', 'Q1', 'Q2', 'Q3', 'Q4', 'Q5', 'Q6', 'Q7', 'Q10', 'Q12', 'Q13', 'Q14') & 7.59(3)&  &\\\cline{1-4} 
('Q0', 'Q1', 'Q2', 'Q3', 'Q4', 'Q5', 'Q6', 'Q7', 'Q10', 'Q12', 'Q13', 'Q14', 'Q16', 'Q19', 'Q22', 'Q25') & 11.9(5) &  &\\\cline{1-4} 
('Q0', 'Q1', 'Q2', 'Q3', 'Q4', 'Q5', 'Q6', 'Q7', 'Q10', 'Q12', 'Q13', 'Q14', 'Q19', 'Q16', 'Q21', 'Q22', 'Q25', 'Q24', 'Q23', 'Q26') & 21.0(7) & &\\\cline{1-4} 
\end{tabular}
    \caption{\textbf{BiRB, DRB, and CRB on IBM Hanoi.} The error rates for every RB experiment we ran on \texttt{ibm\_hanoi}. We ran CRB on up to 5 qubits, we ran DRB on up to 6 qubits, and we ran BiRB on up to 20 qubits.}
    \label{tab:ibm_hanoi_r}
\end{table*}

\begin{table*}[h!]
\begin{tabular}{|l|l|l|l|l|l|l|l|l|}
\cline{1-9}
qubit & $T_1$ (us) & $T_2$ (us) & frequency (GHz)& anharmonicity  (GHz) & readout error & Pr(prep 1, measure 0)& Pr(prep 0, measure 1)& readout length  (ns) \\\cline{1-9}
\Q0 & 170.13 & 240.96 & 5.04 & -0.34 & 0.010 & 0.012 & 0.007 & 817.78 \\\cline{1-9}
\Q1 & 119.38 & 125.05 & 5.16 & -0.34 & 0.013 & 0.013 & 0.013 & 817.78 \\\cline{1-9}
\Q2 & 139.61 & 206.70 & 5.26 & -0.34 & 0.014 & 0.018 & 0.010 & 817.78 \\\cline{1-9}
\Q3 & 120.10 & 32.35 & 5.10 & -0.34 & 0.011 & 0.014 & 0.007 & 817.78 \\\cline{1-9}
\Q4 & 196.74 & 17.02 & 5.07 & -0.34 & 0.006 & 0.006 & 0.007 & 817.78 \\\cline{1-9}
\Q5 & 148.10 & 186.60 & 5.21 & -0.34 & 0.006 & 0.008 & 0.004 & 817.78 \\\cline{1-9}
\Q6 & 97.99 & 143.88 & 5.02 & -0.34 & 0.024 & 0.027 & 0.021 & 817.78 \\\cline{1-9}
\Q7 & 177.40 & 255.79 & 4.92 & -0.35 & 0.012 & 0.014 & 0.010 & 817.78 \\\cline{1-9}
\Q8 & 205.98 & 341.43 & 5.03 & -0.34 & 0.012 & 0.012 & 0.011 & 817.78 \\\cline{1-9}
\Q9 & 96.37 & 208.45 & 4.87 & -0.35 & 0.008 & 0.012 & 0.004 & 817.78 \\\cline{1-9}
\Q10 & 54.74 & 55.26 & 4.82 & -0.35 & 0.020 & 0.021 & 0.020 & 817.78 \\\cline{1-9}
\Q11 & 150.80 & 259.36 & 5.16 & -0.34 & 0.077 & 0.075 & 0.080 & 817.78 \\\cline{1-9}
\Q12 & 96.62 & 175.47 & 4.72 & -0.35 & 0.173 & 0.215 & 0.130 & 817.78 \\\cline{1-9}
\Q13 & 241.95 & 274.17 & 4.96 & -0.34 & 0.047 & 0.045 & 0.050 & 817.78 \\\cline{1-9}
\Q14 & 130.40 & 23.22 & 5.05 & -0.34 & 0.009 & 0.011 & 0.007 & 817.78 \\\cline{1-9}
\Q15 & 80.32 & 35.15 & 4.92 & -0.32 & 0.029 & 0.023 & 0.035 & 817.78 \\\cline{1-9}
\Q16 & 194.51 & 316.64 & 4.88 & -0.35 & 0.009 & 0.010 & 0.008 & 817.78 \\\cline{1-9}
\Q17 & 152.12 & 66.53 & 5.22 & -0.34 & 0.018 & 0.019 & 0.016 & 817.78 \\\cline{1-9}
\Q18 & 155.96 & 138.19 & 4.97 & -0.35 & 0.012 & 0.017 & 0.006 & 817.78 \\\cline{1-9}
\Q19 & 201.17 & 238.68 & 5.00 & -0.35 & 0.006 & 0.007 & 0.004 & 817.78 \\\cline{1-9}
\Q20 & 170.75 & 67.15 & 5.10 & -0.34 & 0.006 & 0.005 & 0.006 & 817.78 \\\cline{1-9}
\Q21 & 128.89 & 31.88 & 4.84 & -0.35 & 0.008 & 0.011 & 0.005 & 817.78 \\\cline{1-9}
\Q22 & 198.10 & 108.40 & 4.92 & -0.35 & 0.011 & 0.012 & 0.010 & 817.78 \\\cline{1-9}
\Q23 & 173.22 & 256.72 & 4.92 & -0.34 & 0.011 & 0.008 & 0.014 & 817.78 \\\cline{1-9}
\Q24 & 158.95 & 36.15 & 4.99 & -0.34 & 0.007 & 0.008 & 0.005 & 817.78 \\\cline{1-9}
\Q25 & 158.51 & 47.02 & 4.81 & -0.35 & 0.010 & 0.010 & 0.009 & 817.78 \\\cline{1-9}
\Q26 & 79.50 & 28.58 & 5.02 & -0.34 & 0.008 & 0.010 & 0.006 & 817.78 \\\cline{1-9}
\end{tabular}
\caption{\textbf{IBM Hanoi calibration data.} Calibration data from \texttt{ibm\_hanoi} from the time of our BiRB demonstrations.}
    \label{tab:ibm_hanoi_calibration}
\end{table*}

\end{document}